\documentclass[fleqn,usenatbib,useAMS]{mnras}

\usepackage{graphicx}	% Including figure files
\usepackage{amsmath}	% Advanced maths commands
\usepackage{amssymb}	% Extra maths symbols
\usepackage{multicol}        % Multi-column entries in tables
\usepackage{diagbox}
\usepackage{bm}		% Bold maths symbols, including upright Greek
\usepackage{pdflscape}	% Landscape pages
\usepackage{journals}

\def\HI{H{\sc i}\, }
\def\HII{H{\sc ii}\, }

\def\kms{$\textrm{km~s$^{-1}$}$}

\usepackage[T1]{fontenc}
\usepackage{ae,aecompl}
\usepackage{newtxtext,newtxmath}
\usepackage{epstopdf}
\title[]{ Arp~58 and Arp~68: two M~51- type systems}
\author[Zasov et al.]{
Anatoly V. Zasov,$^{1,2}$\thanks{E-mail:zasov@sai.msu.ru}
Anna S. Saburova,$^{1}$
Oleg V. Egorov,$^{3,1}$
\newauthor
Vsevolod Yu. Lander,$^{1}$ 
Dmitry I. Makarov$^{4}$
\\
$^1$ Sternberg Astronomical Institute, Moscow M.V. Lomonosov State University, Universitetskij pr., 13,  Moscow, 119234, Russia\\
$^2$ Faculty of Physics, Moscow M.V. Lomonosov State University, Leninskie gory 1,  Moscow, 119991, Russia \\
$^3$ Astronomisches Rechen-Institut, Zentrum f\"ur Astronomie der Universit\"at Heidelberg, M\"onchhofstr.\ 12--14, 69120 Heidelberg, Germany \\
$^4$ Special Astrophysical Observatory, Russian Academy of Sciences, Nizhniy Arkhyz, Karachai-Cherkessian Republic 357147, Russia\\}

\begin{document}
\large%\hfill{``cobs-6.tex'' пїЅпїЅ 17.04.0

\label{firstpage}
\pagerange{\pageref{firstpage}--\pageref{lastpage}} \pubyear{2022}
\maketitle

\begin{abstract}
We study  two M~51-type systems Arp~68 and Arp~58, which strongly differ by their stellar masses, gas content and environment.  Long-slit spectral observations obtained at the 6-m telescope BTA were used to trace the distributions of a line-of-sight (LOS) velocity and a gas-phase oxygen abundance along the spectral cuts. Two systems  are compared by their observed properties. We found a very strong large-scale non-circular motion of gas in both systems and a kpc-size saw-edged velocity profile  along the tidal spiral arm of Arp~68, probably caused by the gas outflow due to the stellar feedback.  A deep decrease of LOS velocity is also found in the `hinge' region in Arp~58, where the inner spiral arm transforms into the tidal spiral arm, which was predicted earlier for M~51-type galaxies. Local sites of star formation and the satellites are compared with the evolutionary models at the colour-colour diagrams.  Unlike the spiral galaxy Arp~58, the main galaxy in Arp~68 system is experiencing an ongoing burst of star formation. Gas-phase metallicity estimates show that Arp~58 has a higher metal abundance and reveals a shallow  negative radial gradient of the gas-phase oxygen abundance. The emission gas in Arp~68  has noticeably lower metallicity than it is  expected for a given  luminosity of this galaxy, which may be connected with its space position in the local void. 

\end{abstract}
\begin{keywords}
galaxies: kinematics and dynamics,
galaxies: evolution
\end{keywords}

\section{Introduction}\label{intro}

M51-type of galaxies present a rather rare variety of interacting spiral galaxies, where a main galaxy has, as a rule, two well-developed spiral arms, and there is  a smaller satellite, observed at the end or near the end of a '' tidal'' spiral arm, being evidently responsible for the formation of the grand design spiral structure of a main galaxy. The first catalog of such systems was compiled by \citet{VV1975} and contains 160 galaxies (including some erroneously classified systems). A more recent catalog by \cite{2008ApSS.315..249J} already lists 232 systems of M~51-type. Most of such systems contain spiral galaxies of late morphological types, characterized by the enhanced star formation rate \citep[see e.g.][]{KlimanovReshetnikov2001}. The kinematics of galaxies of this type is characterized, as a rule, by the presence of noticeable non-circular gas motion, while the radial distribution of their line-of-sight (LOS) velocity is often asymmetric, though a strong distortion of rotation curves is observed only in a few cases \citep{Klimanovetal2002,2016AJ....152..150G}. In some systems, deep local minima are observed in the LOS distribution curve, the cause of which remains unclear (as an example, among the 12 binary systems studied by \citealt{Klimanovetal2002}, such peculiarities are noticeable in NGC~151, NGC~7753 and NGC~7757). 

M51-type galaxies are the systems we observe  at a certain stage of evolution, preceding to the merging, when the rearrangement of the spiral structure, mixing and radial movement of gas, which are reflected in their observed characteristics, become significant.

Using   the kinematic sample of 21 interacting M51-type galaxy systems \cite{2016AJ....152..150G} found that the radial
velocity distributions present asymmetries in about half of the studied main galaxies. Some of these galaxies also have local extrema of their rotation curve, however there is only one object (AM 1304-333) where the rotation curve is highly distorted. As it was argued by \cite{2016AJ....152..150G}, the
orbital motion of the satellite is within the range of
amplitudes of the rotation curve of the main galaxy which agrees with the natural expectation that the orbital plane of a satellite is close to the main galaxy' disc plane.

Numerical simulations confirm that the tidal forces induced even by a low-mass satellite under favorable conditions may reproduce the observed regular spiral structure of M~51-type galaxies. In agreement with observations, they show that the perturbed gas motion may have a very complicated character which differs from that expected for the stationary density wave, moving through a disc with the single global pattern speed \citep[see e.g.][and references therein]{Dobbs2010, Pettitt2017}. However a general picture of gas motion at a kpc-scale, a distribution of gas-phase metallicity, and the  connection of velocity perturbation of gas with the observed sites of star formation are poorly known, being studied in details only for a few galaxies of M51-type. 

Previous long-slit studies of small samples (less than 20) of interacting galaxies revealed more shallow radial metallicity gradient than in isolated spiral galaxies \citep[e.g.][]{Rupke2010, Rosa2014}. Large integral-field unit (IFU) surveys allowed to investigate the spatially resolved properties of the interacting galaxies for significantly larger samples. It was shown that massive interacting galaxies exhibit kinematic asymmetries \citep{Bloom2018, Feng2020} and enhanced star formation rate \citep{Barrera-Ballesteros2015, Pan2019}. Shallow oxygen abundance gradient was also confirmed \citep{Sanchez2014}, however this is possibly applicable  only to those galaxies which are highly distorted by interaction, while several galaxies chosen just by the evidence of some interactions do not differ by their oxygen abundance gradient (expressed as [dex$/r_e$])  from the samples of 
 isolated or paired galaxies  \citep{Sanchez-Menguiano2018}. 
 
 These IFU studies  of large samples of interacting galaxies mentioned above are focused on the scales exceeding 1~kpc, while the number of systems with available 2D distribution of the ionized gas kinematics and/or metals abundances at the small scales is still scarce \citep[see, e.g.,][]{MunozElgueta2018, Karera2022}, and the long-slit spectroscopy remains a major source of information for such objects.

 The present work continues a series of studies of the emission gas and star formation regions on the periphery of interacting galaxies and between them, carried out at the BTA using a long-slit spectroscopy \citep{zasovetal2015, zasovetal2016, Zasovetal2017, Zasovetal2018, Zasov2019, zasovetal2020}. The main emphasis in these papers was aimed on the features of gas kinematics, the excitation mechanism of gas emission, the distribution of gas-phase metal  abundance, and the ages of regions of star formation in the systems. 

 Here we consider two interacting systems of galaxies of M~51-type  (Arp 68 and Arp 58) which vary greatly  in  their mass and luminosity  in spite of a similar morphology. 

One of the system  (Arp 58) was observed by us earlier, although the only spectral cut we obtained turned out to be insufficient for confident conclusions regarding the kinematic properties of the galaxy and a gas-phase metallicity distribution in the tidal spiral arm in this galaxy. Below we give a short description of these systems.

\begin{figure}
\includegraphics[width=\linewidth]{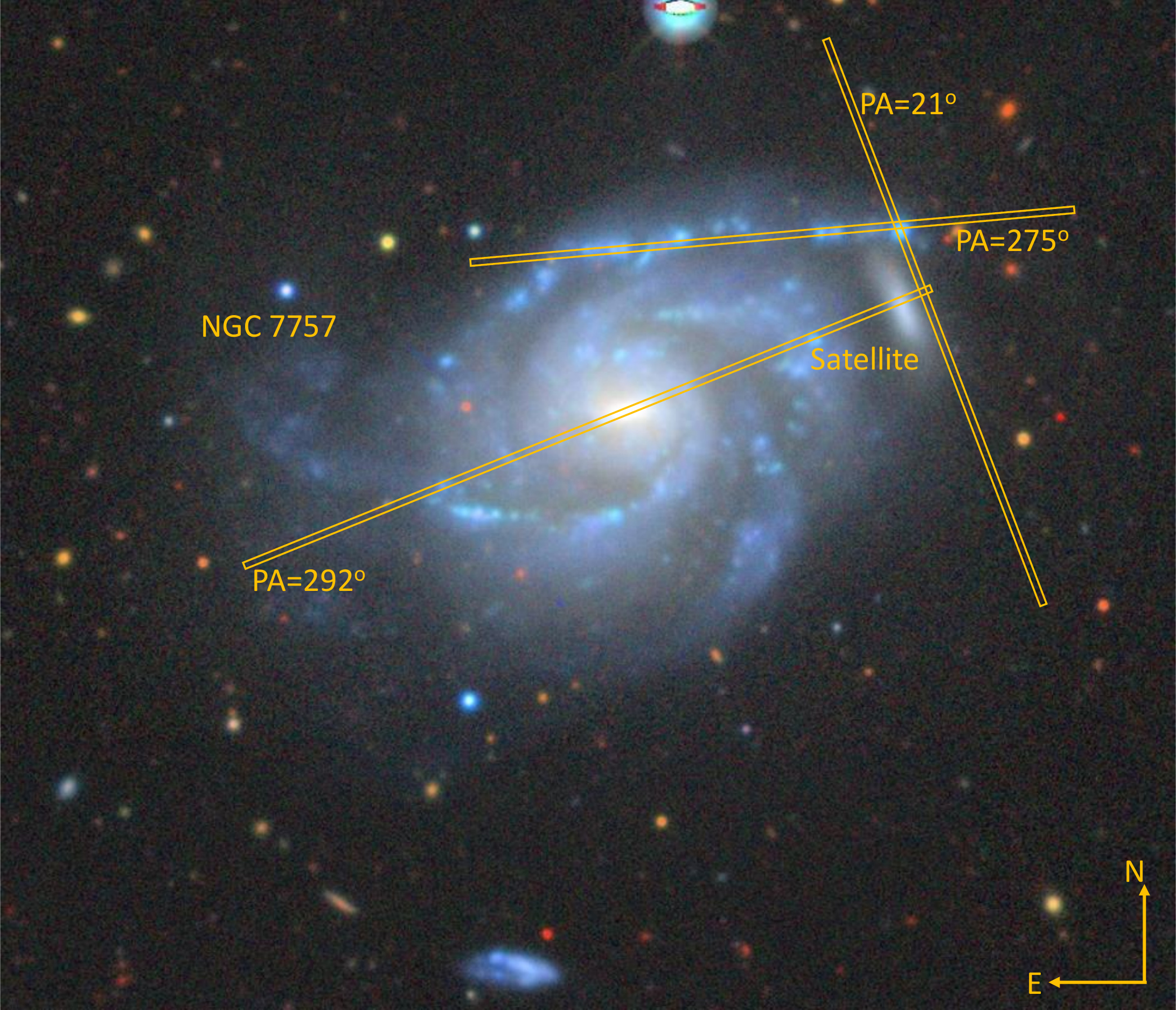}
\caption{Composite {\it g,r,z}-band DECaLS image of Arp~68 with the overplotted positions of the slits used for spectral observations in current paper.}
\label{map}
\end{figure}

The system Arp~68 (or VV 407) includes a spiral galaxy  NGC~7757 and its small featureless elongated satellite observed near the end of the most prominent spiral arm (see Fig \ref{map}), which allows us to ascribe this system as the M~51-type.
   
Distance estimates for this galaxy are contradictory \citep[see the Hyperleda database,\footnote{http://leda.univ-lyon1.fr/}  ][ which  we use for  the basic parameters of the galaxy]{Makarov2014}.  Indeed, the distance value based on the galaxy systemic velocity  V $\approx$ 3000~\kms{} (corrected for the Virgo flow) is 43 Mpc for the adopted $H_0 = 70$~\kms{}\,Mpc$^{-1}$, which corresponds  to the B-band total absolute magnitude $M_B = -20.35$, or the total luminosity $2.15\cdot10^{10}L_\odot$.  At the same time, the rotational velocity of this galaxy obtained from the \HI profile (which, by the way,  is rather symmetric)  is 93~\kms{} for the adopted inclination angle of the disc   $i= 44^\circ$ (Hyperleda). Being applied to the Tully-Fisher relationships, it gives   $D = 15.8$~Mpc (Hyperleda) which leads to the almost  an order of magnitude lower luminosity $L \approx 3\cdot10^{9}L_\odot$. The  systemic cosmological velocity expected for this  distance should be about 1100~\kms, which is lower at nearly 2000~\kms{} than the observed value. So this low distance estimate is very questionable,  especially if to take into account that the interacting galaxies badly follow the classical Tully–Fisher relations  (see \cite{2016AJ....152..150G}).  In addition, a well defined spiral structure of this galaxy and rather thin and contrast spiral arms  is difficult to reconcile with the low luminosity of the galaxy. By these reasons  we will use the distance $D = 43$~Mpc below.

LOS velocity distribution along the major axis obtained from the measurements of emission lines by \citet{Klimanovetal2002} revealed non-symmetric velocity profile along the radial distance and  strong non-circular velocities in some local regions, which  does not allow to estimate reliably  the velocity of rotation. The authors report the upper limit of the rotational velocity (non-corrected for the galaxy inclination) to be 100~\kms as obtained for the eastern half of the disc, and significantly lower value if measuring at the opposite side.

  Arp 68 was reported to have an integrated \HI flux of 14.89 Jy \kms ~from ALFALFA \citep{Haynes2018} and 15.3 Jy \kms ~from the previous HIPASS survey \citep{wong2006}. This yields $M_{\rm HI} \sim 6.5$ to $6.7\cdot 10^9 M\odot$ for the assumed distance. The ratio of the gas mass to the B-band luminosity  $M_{\rm HI}/L_B \approx 0.3$ 
 evidences that this is a gas-rich galaxy (note that this ratio is invariant to the accepted distance).
\begin{figure}
\includegraphics[width=\linewidth]{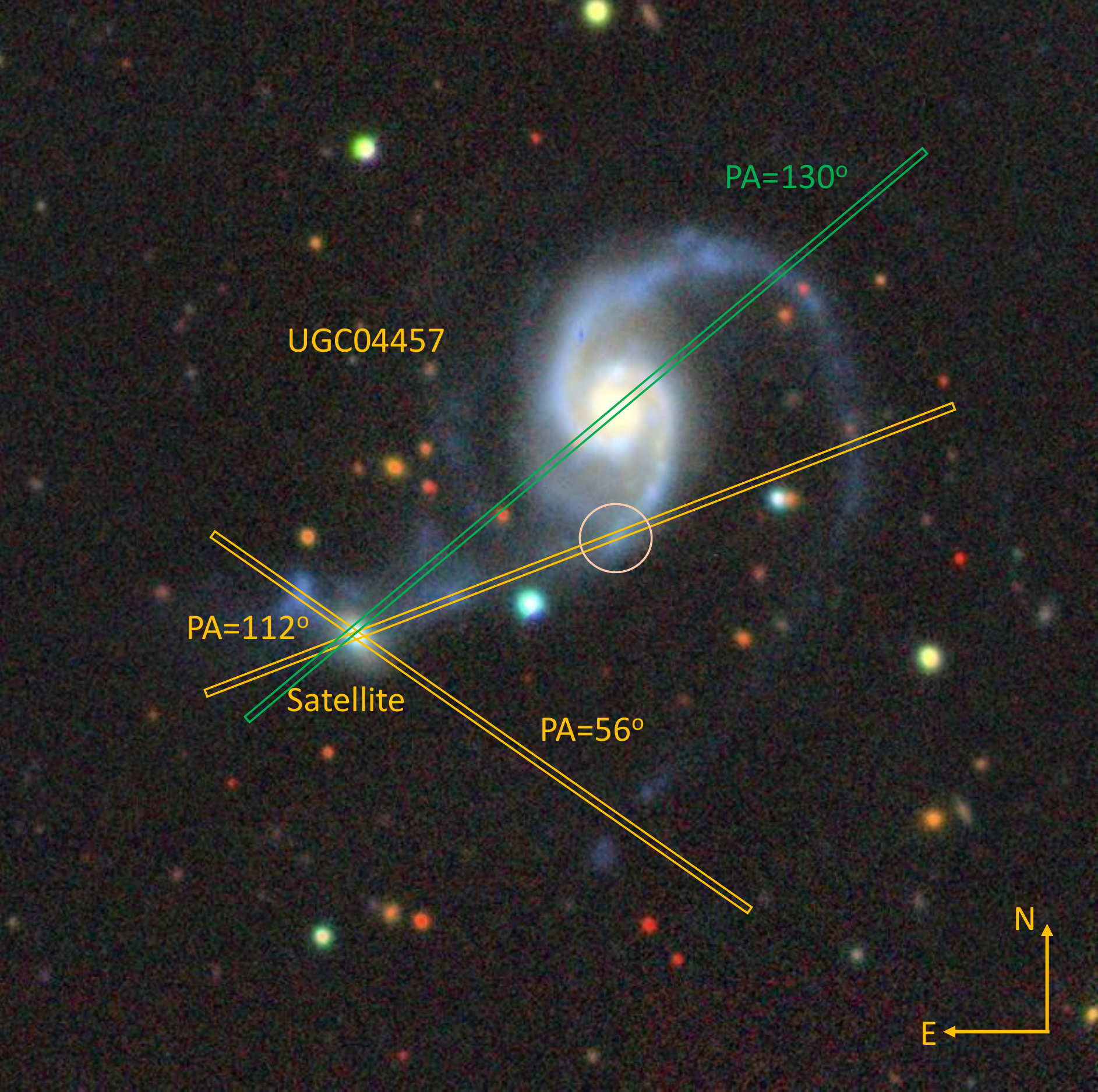}
\caption{The same as in Fig. \ref{map}, but for  Arp~58. Green symbols show the slit position from Paper I.The circle gives the position of the 'hinge' region discussed in the text.}
\label{map2}
\end{figure}

The main spiral galaxy  in Arp~68 system  is characterized by the active star formation, which follows from its low colour indices: $(B-V)_c = 0.32^m$, $(U-B)_c = -0.31^m$ taken from Hyperleda. Judging from the distribution of starforming regions, they are mostly concentrated in the main  spiral arms which possess a clumpy structure, especially the arm extended towards the satellite,  which contains the brightest clumps.  Note that  this galaxy is comparable with M~33 in the Local Group by the luminosity and gas richness  \citep{corbelli2003}. However, unlike the latter, it has a well-developed spiral pattern (Grand Design). Its formation is apparently due to tidal action from a nearby satellite, rotating in the orbit in the same direction as the main galaxy.

Another galaxy discussed in this paper is  Arp~58 = UGC 4457 = VV 313. This is a highly luminous  bulgeless galaxy that has two very open spiral arms, one of which  is tidally extended toward the compact satellite observed near the end of  the arm (at least in a projection).  A systemic velocity of the spiral galaxy is about 11000~\kms. 
 
 In the previous paper \citep{Zasov2019} (hereinafter -- Paper I), we gave more detailed description of this galaxy. The results of spectral observation of Arp~58 given in Paper I were based only on a single spectral cut obtained at the 6-m telescope BTA, where the slit passed through the centers of the main galaxy and the satellite. In particular, it was found that the main galaxy, unlike the satellite, reveals a regular rotation, whereas the LOS velocity profile of the satellite  is highly asymmetric, and the  ionized gas  at its periphery is disturbed and revealing the elevated velocity dispersion in comparison with the inner region. A rough estimate of the galaxy's rotation velocity ($\sim$ 230~\kms{} after correction for the disc inclination) confirmed its large mass within the  B-band isophotal diameter $D_{25}$ (around $10^{11}M_\odot$). The authors pointed out a strange blue ejection-like feature that appears bright in UV light (GALEX). Visually it emanates from the satellite body and it was proposed to be a continuation of the distorted tidal spiral arm behind the satellite. The oxygen abundance gradient along the radial distance of the central galaxy is small, being in the range  0.005--0.01 dex kpc$^{-1}$.

 In this paper we study the dynamics, gas-phase oxygen abundance and star formation features of these two M~51-type galaxies, which strongly differ by their luminosity and mass. 

 The paper is organized in the following way. In Section \ref{obs} we give the details on observations and data reduction. The results of observations are presented in Sections \ref{res1} for Arp~68  and  Section \ref{res2} for Arp~58. Section \ref{sec:disc} gives the discussion, and the last Section  \ref{conclusions} contains a short summary of the main results.

\section{Observations and data reduction }\label{obs}
We carried out  optical long-slit spectral observations of Arp~68 and Arp~58 at the Russian 6-m BTA telescope  of Special Astrophysical Observatory with the focal reducers SCORPIO-1 \citep{AfanasievMoiseev2005} and SCORPIO-2 \citep{AfanasievMoiseev2011}.  The log of observations is given in Table~\ref{obs}. 
We used three slit positions for Arp~68 and two positions for Arp~58 (in addition to the spectral cut used in Paper I for the latter galaxy). The Legacy Survey DR9 \citep{Dey2019} composite {\it g,r,z}-band images of the systems  with the positions of the slits are shown in Figs.\ref{map}, \ref{map2}. For Arp~68 the position angle (PA) of the first cut PA= $292^\circ$ is close to the major axis PA = $120^\circ$, so the difference of the LOS velocities of circularly moving gas due to this discrepancy might be small, not exceeding  $\sim$2 per cent. 

A procedure of data reduction was described in details in our previous papers \citep[see e.g.,][]{zasovetal2015,zasovetal2016,Zasovetal2017}. Shortly, it includes a bias subtraction and truncation of overscan regions, flat-field correction, the wavelength calibration based on the spectrum of He-Ne-Ar lamp, cosmic ray hits removal, summation of individual exposures, and the night sky subtraction.  The instrumental profile of the spectrograph was determined by the fitting of the twilight sky spectrum observed in the same observation run. The reduced  spectra of the galaxies were fitted with high-resolution PEGASE. HR \citep{LeBorgneetal2004} simple stellar population (SSP) models using the NBURSTS full spectral fitting technique \citep{Chilingarian2007a, Chilingarian2007b}. After subtraction of the model stellar spectra we  fitted the emission lines by Gaussian profiles which enabled us to get the velocity, velocity dispersion of the ionized gas and the fluxes of emission lines.

\section{Arp~68}\label{res1}
 
\subsection{Kinematics of the ionized gas} \label{res}
\begin{figure}
\includegraphics[width=\linewidth]{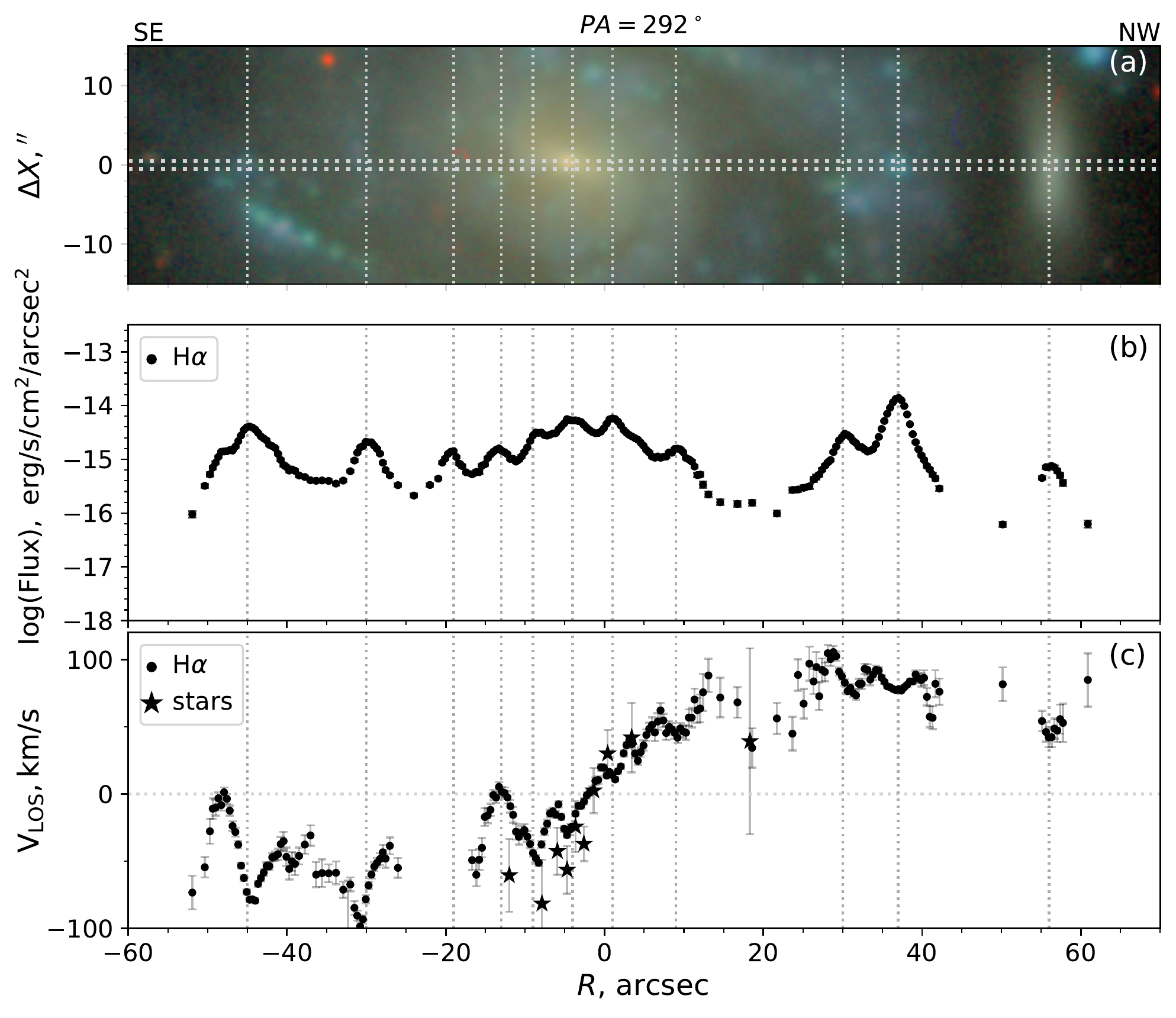}%11.5cm
\caption{The reference {\it g,r,z}-band image of Arp~68 (a). The radial profiles of the $H_\alpha$ flux (b) and  line-of-sight velocity of gas and stars relative to the systemic velocity $V=2981$~\kms{} (c) along PA = 292$^\circ$.}
\label{arp68_pa112}
\end{figure}

\begin{figure}
%\hspace{-4.5cm}
\includegraphics[width=\linewidth]{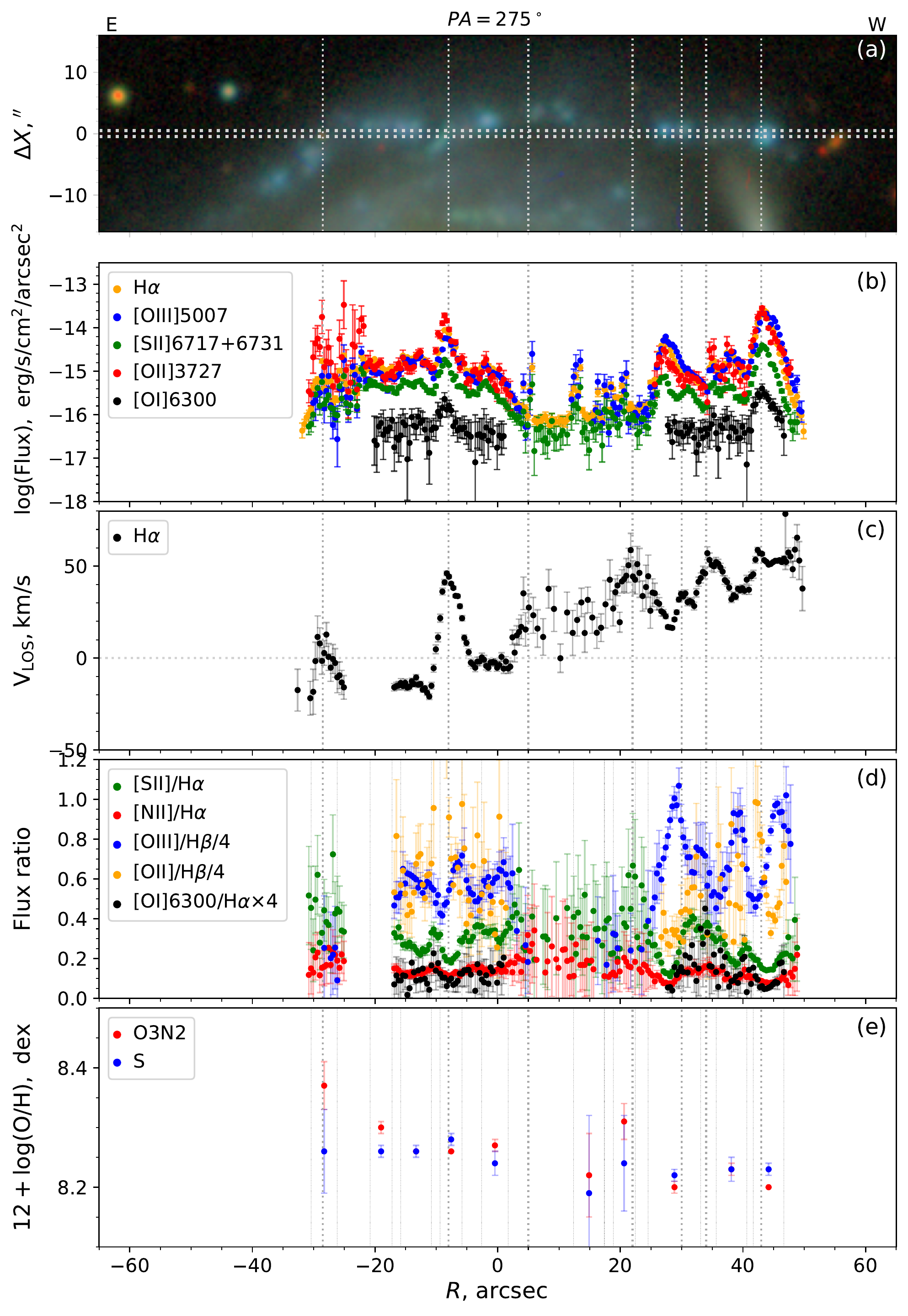}%11.5cm
\caption{The reference {\it g,r,z}- band image of Arp~68 (a). The radial profiles of the emission lines fluxes (b), line-of-sight velocity  relative to the systemic velocity which we found to be  $V=2981$~\kms{} (c), the emission lines fluxes ratios (d), the oxygen abundance obtained by O3N2 and S-methods (e) along PA = 275$^\circ$.}
\label{arp68_pa95}
\end{figure}

\begin{figure}

\includegraphics[width=\linewidth]{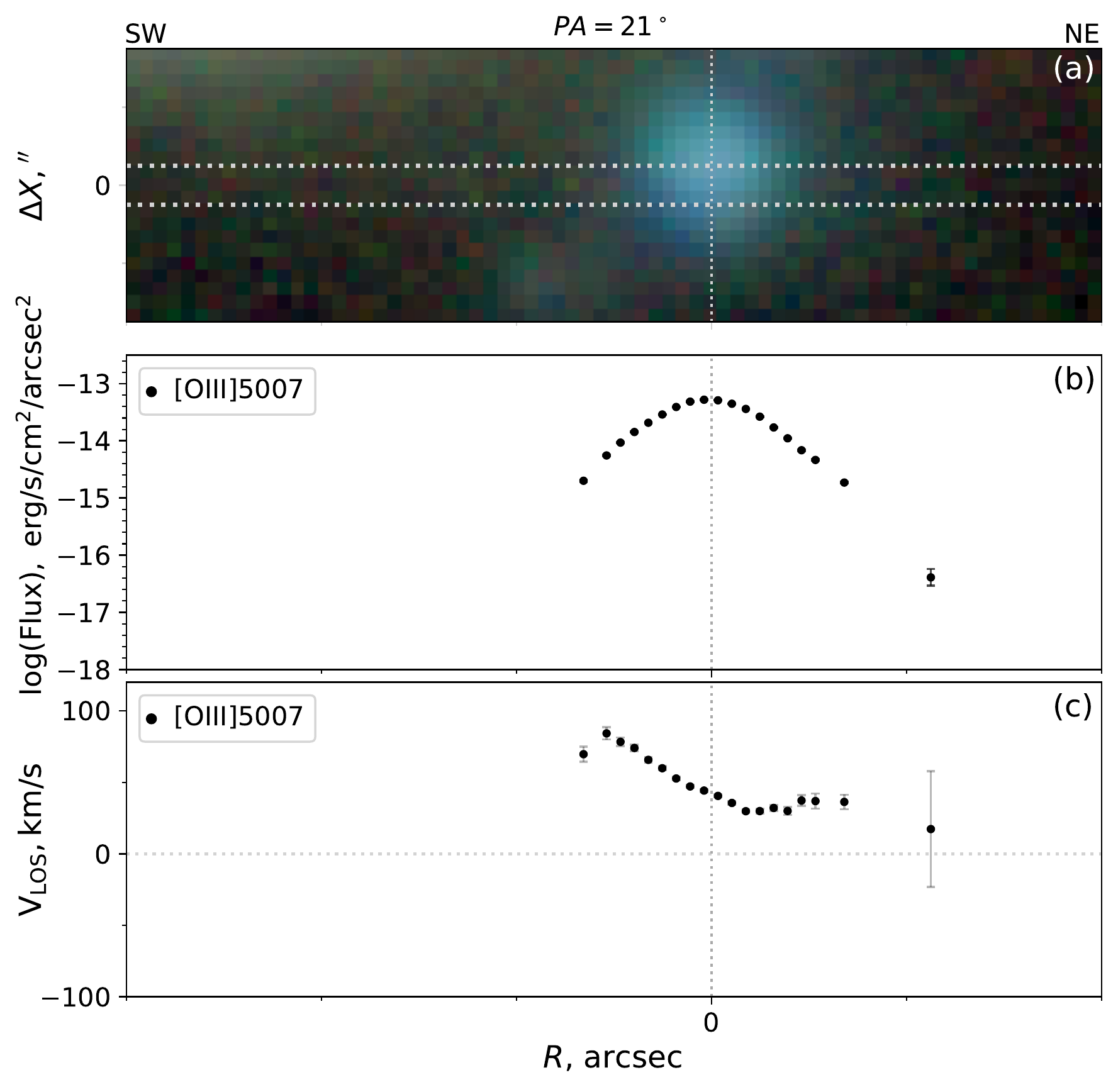}
\caption{The same as in Fig. \ref{arp68_pa112}, but for the slit crossing the star-forming clump in the spiral arm of Arp~68 ($PA=21^\circ$).}
\label{arp68_pa202}
\end{figure}

The distributions of LOS gas velocity of Arp~68 along the slits are shown in Figs. \ref{arp68_pa112}, \ref{arp68_pa95}  and \ref{arp68_pa202}. 
A general trend of LOS velocity reflects the rotation of the galaxy where the western part of  its disc recedes from the observer. Velocity profile along the slit PA=$292^\circ$  (Fig.\ref{arp68_pa112})  shows that the LOS velocity range along the major axis is about 180~\kms.  

LOS velocity of the satellite (the point group on the far right of the velocity distribution in Fig. \ref{arp68_pa112}) is about 50~\kms{} higher than the central velocity of the main galaxy and about 30~\kms{} lower than the velocity of the adjacent area of the disc. It nicely agrees with the self-evident expectation that a satellite which orbits in the same direction as a disc of a galaxy has the strongest tidal effect on its internal structure. 

A general shape of the velocity profile along the major axis is asymmetric with respect to the photometric centre,  as it  was earlier pointed by  \citet{Klimanovetal2002}: the approaching half of the disc looks like it's rotating at about 30~\kms{} slower  than the receding one, although the $r$-band SDSS image photometry shows that the optical centre of the disc coincides with the kinematic one within 1--2~arcsec.

In addition, there are deep local velocity minima at the velocity profile evidencing a strong non-circular motion of emission gas at kpc scale. The largest one is at $r = 15$--25 arcsec to the west of nucleus between the  spiral arms.  However, it is difficult to say how much the gas velocity in this region differs from the circular velocity, since the circular rotation curve of this galaxy is determined rather roughly. 2D spectroscopy is definitely required here.  It is worth noting that  a strong local LOS velocity variations of gas are also observed in some other systems of  M~51-type:  see, for example, NGC~7753 \citep{Klimanovetal2002}, NGC~3227 \citep{Keel1996}, or the system Arp~58 (see below).

The stellar  and ionized gas LOS velocities (black stars and circles in Fig \ref{arp68_pa112}) we managed to obtain reveal  the  asymmetric profile. The most strong disturbances appear for ionized gas. Less perturbed stellar rotation agrees with the   absence of significant asymmetry of r-isophotes between N and S-halves of the galaxy, although there is a variation of photometrical estimates of orientation angles of the disc along the radius in the inner disc (see Fig. \ref{arp68_phot}), which may be related  to the bar and/or a prolate bulge of the galaxy.

\begin{table*}
\caption{Log of observations}\label{log}
\begin{center}
\begin{tabular}{cccccccc}
\hline\hline
& PA & Date & Exposure time& Grism&Spectral range&Dispersion& Seeing\\
 &  ($^\circ$)  &  &     (s) & & \AA&\AA/pix&      (arcsec) \\
\hline
Arp~68&21&15.11.2020&1200&VPHG1200B&	3600--5400&0.87&1.6 \\
Arp~68&275&15.11.2020&2700&VPHG1200B&	3600--5400&0.87&1.7 \\
Arp~68&275&15.11.2020&1200&VPHG1200R&5700--7500&0.86&1.7 \\
Arp~68&292&15.11.2020&1200&VPHG1200R&5700--7500&0.86&1.7 \\
Arp~58&112&9.12.2020&5400&VPHG1200@540&3650--7250&0.87&1.7\\
Arp~58&56&11.12.2020&5400&VPHG1200@540&3650--7250&0.87&1.7\\
\hline

\end{tabular}
\end{center}
\end{table*}

The most peculiar  behaviour  of the velocity  of ionized gas in this galaxy is the variation  of LOS velocity on kpc- scales, which gives the velocity profile a saw-tooth shape for both spectral cuts, especially   for the slit running along the arm PA=$275^\circ$ (Fig. \ref{arp68_pa95}). The LOS velocity varies at 20--40~\kms{} on a scale of about 10 arcsec, which corresponds to 2 kpc. Due to a projection effect,   the amplitude of the velocity variations should be even higher if the  perturbations lie in the disc plane. Velocity profile for the slit passing through the centre of the galaxy (PA=$292^\circ$) also shows small-scale variations, however in general it looks more blurry. Nevertheless there are some regions, where the velocities fall down to zero (the approaching side) or to nearly zero (receding side) with respect to the central value (see Fig. \ref{arp68_pa112}). Significant decreasing of LOS velocities are observed between spiral arms in the inner part of the galaxy without a clear connection with  any   morphological details.

Small-scale irregularities in the LOS  velocity along the northern spiral arm (PA = 275$^\circ$) are definitely connected with the bright \HII- regions. Indeed, a comparison of brightness and velocity distributions shows that, in almost all cases, local velocity minima occur in the \HII region (see  Figs.   \ref{arp68_pa112}, \ref{arp68_pa95} where the vertical lines   connect the emission regions with the  corresponding sections of the velocity and intensity profiles).  The exception is  the velocity  maximum at $R = -8$~arcsec  (Fig. \ref{arp68_pa95}) which also falls on the region of increased line emission. However in this case the slit crosses only the outer part of the extended  \HII region, so the velocity maximum   may have another origin.   

The observed  effect may hardly be associated with the variation of gas velocity caused by the interaction of gas with the spiral density wave, since the velocity jumps have the same sign on both sides of the minor axis, which implies that  its origin is not related with the rotation of the disc. Most probably we observe here  the outflow of gas  due to stellar feedback  from the regions with the high concentration of young stars (a stellar wind, or the expanding gas shells) in the presence of an absorbing medium, when we see only the emission gas ejected towards the observer.

 The velocity distribution across the brightest complex of the emission  at the very tip of the northern spiral arm (PA=275$^\circ$) is the most surprising. The   diameter of its bright inner part is about 5--7 arcsec, or 1--1.5  kpc for the adopted distance. The spectral cut (PA=275$^\circ$)  does not reveal significant velocity gradient, while  the near perpendicular cut  PA=21$^\circ$ (Fig. \ref{arp68_pa202}), passing through the clump, clearly shows a strong variation of LOS velocity at about 40~\kms{} across the clump. If we interpret this velocity gradient as a result of stellar feedback, then we should  assume the outflow  of gas within a wide cone at a velocity of 30--40~\kms{} in the direction of the observer. The alternative scenario explaining the observed gas kinematics in this clump could be a disc rotating at a speed of about 20~\kms. In this case the dynamic mass of the clump should be at least $10^8M_\odot$, and the rotation period of about $10^8$ years, which excludes its formation as a relaxed massive system in a  short  time interval during the phase of a strong tidal interaction between galaxies. In addition, the axis of its rotation should not be directed perpendicular to the disc, but approximately along the spiral, otherwise we would observe a  noticeable gradient in this direction. 

\subsection{Gas-phase oxygen abundance in Arp~68}\label{sec:Arp68_abund}

\begin{figure*}
%\hspace{-4.5cm}
\includegraphics[width=\linewidth]{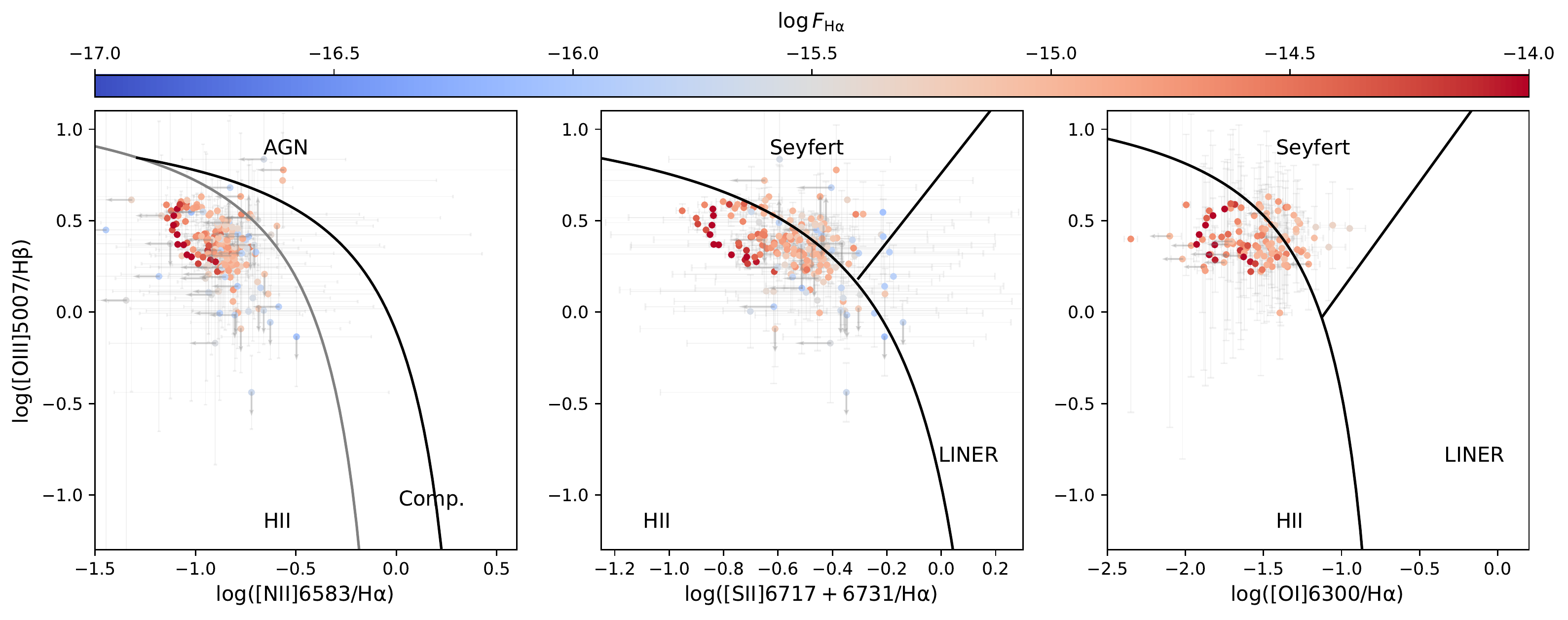}%11.5cm
\caption{The diagnostic BPT diagrams for individual bins along the slit $PA=275\degr$ for Arp~68. Colour corresponds to the measured surface brightness of H$\alpha$ line. The regions above the \citet{Kewley2001} demarcation line (black curved line) are unlikely ionized by massive stars only. Composite mechanism of gas excitation is expected for the regions lying between that line and \citet{kauffmann03} line (gray) on the left-hand panel. Straight black line (from \citealt{Kewley2006}) on the central and right-hand panels  separates the domains typically occupied by Seyferts and low-ionization narrow emission-line regions (LINERs)}
\label{arp68_bpt}
\end{figure*}

In Fig. \ref{arp68_bpt} we present the diagnostic BPT diagrams \citep{BPT, BPT_S} constructed for individual bins along the slits, which give the information about possible mechanisms of excitation of atoms. In most cases we are dealing with  the photoionization by massive stars typical for \HII regions. However some regions reside above the black curve (maximum starburst line from \citealt{Kewley2001}) evidencing the additional mechanisms of excitation due to the presence of diffuse ionized gas (DIG) or  shock waves, caused by SN remnants. These regions are in most cases associated with the low intensity regions. 
We excluded them  from the further evaluation of gas metallicity.

We measured the oxygen abundance $\mathrm{12+\log(O/H)}$ in the \HII regions distributed along the spiral arm (PA=275$^\circ$). For that purpose we integrated the emission spectrum within the borders shown by vertical solid lines in panel (e) of Fig.~\ref{arp68_pa95} for each region and measured the fluxes of the bright emission lines (H$\alpha$, H$\beta$, [O~\textsc{iii}] 4959, 5007~\AA, [N~\textsc{ii}] 6548, 6584~\AA, [S~\textsc{ii}] 6717, 6731~\AA). Line ratios were corrected for reddening using the Balmer decrement. From the ratios of the measured fluxes we estimated the oxygen abundance  using two empirical calibrations: S-method \cite{Pilyugin16}, and
O3N2 method \citep{Marino2013}, although the latter one is at the limit of its  applicability for a given metallicity. Oxygen abundance $\mathrm{12+\log(O/H)}$ slowly falls down along the spiral arm remaining between 8.2 and 8.3, which is about three times lower than the solar abundance. After taking into account the orientation angles of the disc, it roughly corresponds to radial gradient of about  $0.01$~dex~kpc$^{-1}$ though the radial extent is too small for precise measurements. A low abundance gradient is typical for tidal structures, most probably evidencing the radial motion of gas. 

The measured oxygen abundance of \HII regions in Arp 68 appears to be too low for such a luminous  galaxy. It is rather typical for galaxies with $M_B \approx -17^m.. -18^m$  \citep[see e.g.][]{Berg2012}, while the galaxy under discussion is two or three  magnitudes brighter.  There is no indication that this anomaly is the result of interaction with the satellite. It is important, however, that Arp 68 is located inside the nearby void in Eridanus. The metal deficiency of gas was found for low luminous galaxies of this void  earlier by \cite{Kniazev2018}. They compiled the available data on O/H ratio in 36 low luminous void galaxies and compared them with the control sample of galaxies of similar type and luminosity, revealing a clear evidence for a lower average metallicity of the Eridanus void galaxies. Earlier the reduced metallicity was found  in galaxies in some other voids  \citep[see][ and references therein]{Pustilnik2021}. 

It is worth noting that the main galaxy  in Arp 68 is a mildly  luminous spiral, not a dwarf galaxy, although it is unusually rich of gas, which makes it similar to late-type dwarf galaxies. This abundance anomaly is naturally treated as indication of slower evolution of low-massive galaxies in  voids with respect to similar objects residing in denser environment  \cite{Kniazev2018}.

\subsection{Stellar mass distribution in NGC~7757 (Arp~68)}
\begin{figure*}
\hspace{-1.5cm}
\includegraphics[height=0.4\linewidth, trim={0 1.2cm 0 0}]{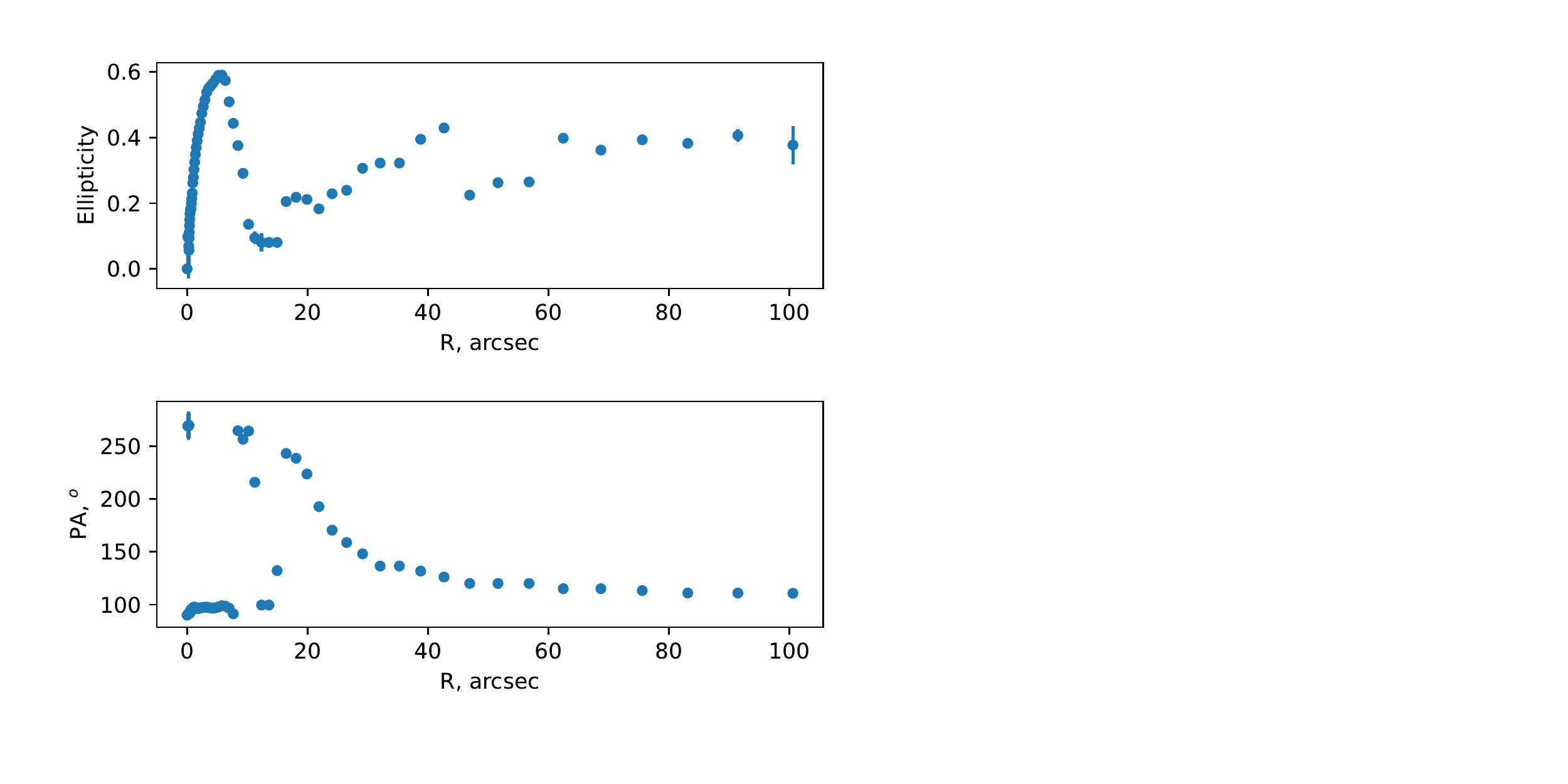}%11.5cm
\vspace{-0.0cm}
\hspace{-7.0cm}
\includegraphics[height=0.4\linewidth]{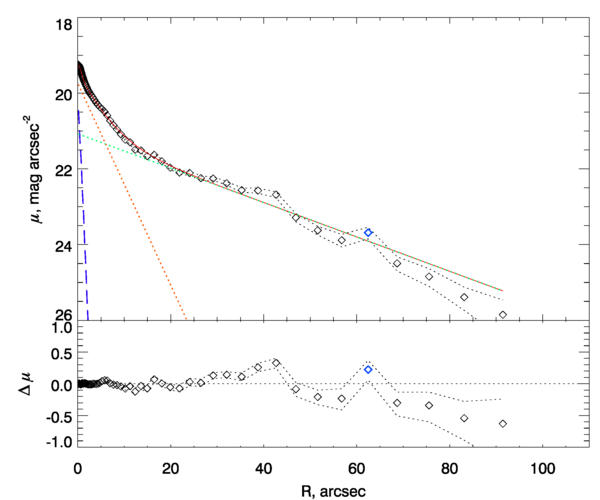}%11.5cm

\caption{The results of the isophotal analysis of the {\it r}-band image of Arp~68. The azimuthally averaged radial profiles of ellipticity (left-hand top panel), positional angle of the photometric major axis (left-hand bottom panel) and the surface brightness profile (right-hand panels) decomposed into the  Sersic  component (blue dashed line) and two exponential discs (red and green dotted lines). The residuals are shown in the bottom panel. }
\label{arp68_phot}
\end{figure*}

We performed isophote analysis of DECaLS {\it r}-band image of the spiral galaxy in Arp 68 using the {\sc ellipse} task in the {\sc PHOTUTILS python} library \citep{photutils}. The resulting radial variation of the ellipticity, position angle and {\it r}-band surface brightness is shown in Fig. \ref{arp68_phot}. The photometric r-band profile  of the galaxy is rather smooth  and does not reveal any significant distortion of the stellar disc at least within the radial distance between 30 and 100 arcsec ( see Fig. \ref{arp68_phot}). 

The surface brightness profile  may be decomposed into three components: a Sersic bulge and two exponential discs with different scalelengths.  For the Sersic component the  effective radius  $R_{\rm eff} = 0.68 \pm 0.80$  arcsec,   central surface brightness  $\mu_0 = 20.45 \pm 1.87$ mag/arcsec$^2$, Sersic index $n_{\rm ser} = 0.88\pm 0.21$.  
The exponential discs are characterized by the scalelengths $(R_{\rm exp})_1 =  23.92 \pm   1.15$ arcsec, $(R_{\rm exp})_2 =   4.07\pm   0.19$ arcsec, and central surface brightnesses $ (\mu_0)_1 =  21.07 \pm   0.07$ mag/arcsec$^2$,  $(\mu_0)_2 =  19.74\pm   0.02$ mag/arcsec$^2$.

Using the photometric data, it is possible to estimate the disc contribution  to the total amplitude of the rotation velocity of the galaxy assuming the position and inclination angle obtained from the photometry: $PA=120^\circ$, $i=42^\circ$.  To move to the stellar surface density of the disc we used the $(g-r)$-colour obtained from the DECaLS images of Arp~68 and M/L-colour relations from \citet{Roediger2015}. 

\begin{figure}
%\hspace{-4.5cm}
\includegraphics[width=\linewidth]{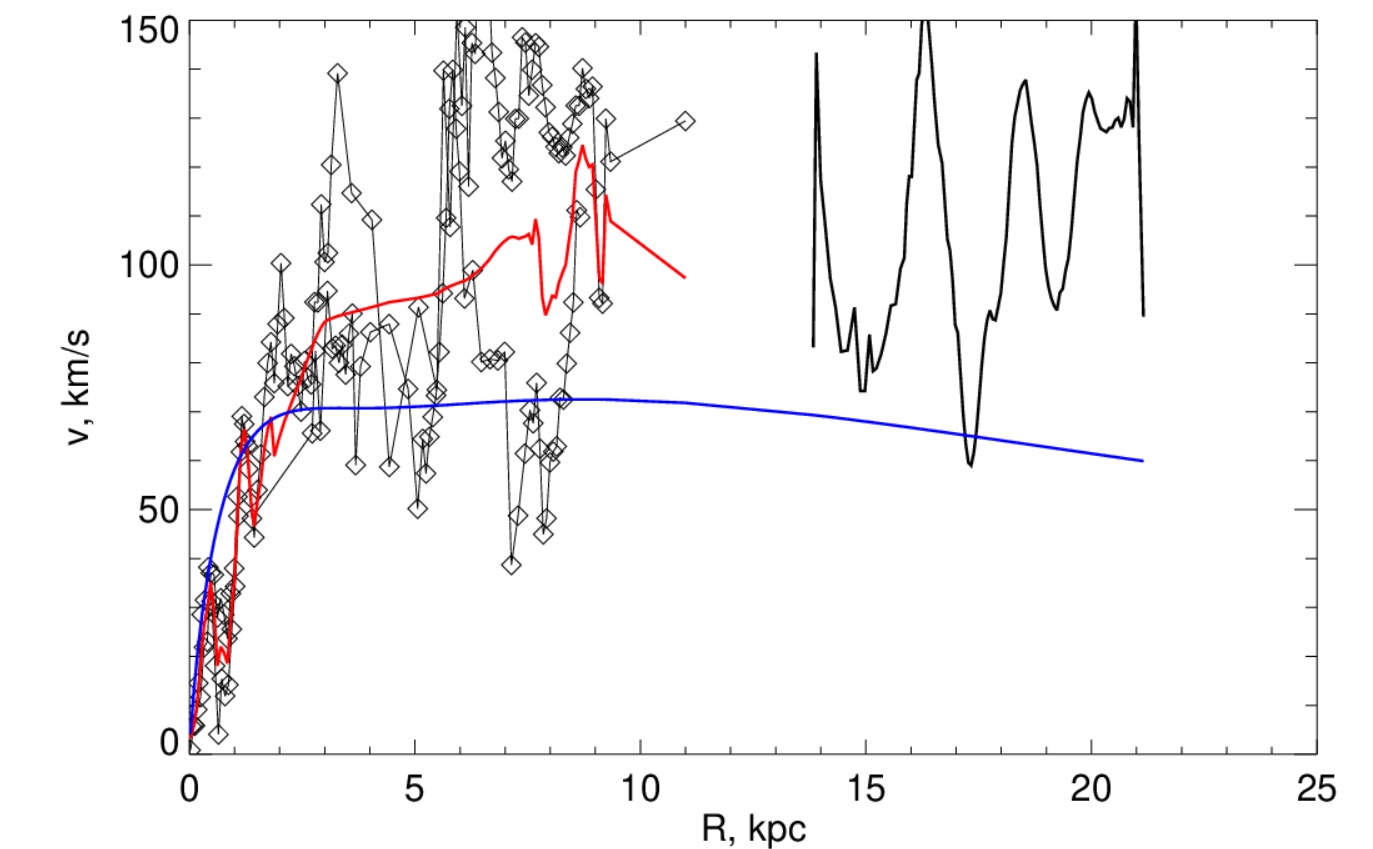}%11.5cm
\caption{Black lines denote the formally constructed indicative rotation curve of Arp~68, based on three spectral cuts, which reveals strong non-circular velocities of emission gas and the LOS asymmetry. A photometry-based disc contribution to the rotation curve is marked  by the blue line. Red line shows the smoothed arithmetic mean between the two sides for the central region.}
\label{arp68_rc}
\end{figure}

The rotation curve which corresponds to its photometric profile (without a dark halo) is drawn by the blue solid line in Fig. \ref{arp68_rc} to compare it with the  formally calculated rotation curve obtained from the gas velocity measurements for two halves of the galaxy,  conventionally assuming a circular rotation for the gas. One can see from Fig. \ref{arp68_rc}, that the strong non-circular motions hinder mass-modelling of the galaxy on the base of gas kinematics. As it was mentioned above,  one side of Arp~68 looks as if it rotates  slower than the other one. Note that the non-corrected half-widths of \HI line of the galaxy at 50 per cent level of the maximum \citep[75~\kms{} according to][]{Haynes2018} after correction for disc inclination gives maximum  of rotation velocity of about 110~\kms{} which better  agrees with the \HII data for the western half of the disc with a higher velocity of rotation.  Maximal velocity $\sim 100$~\kms{} also agrees  better with the baryonic mass estimate according to the baryonic Tully-Fisher relation  \citep[see e.g.][]{Ponomareva2018}, although the application of this dependence for strongly interacting galaxies is not entirely justified.

As it follows from Fig. \ref{arp68_rc}, the photometrically based estimate of disc contribution to the rotation curve (blue line) agrees  with the version of higher velocity of  disc rotation  ($\sim 100$~\kms)  if to take into account that in the moderately slow rotating  galaxies the observed velocity of rotation is usually higher than the baryon component of rotation curve due to the presence of dark halo \citep[a good example is M33 with a high contribution of dark halo in the rotation curve ][which has a comparable  luminosity to Arp68 and the maximal velocity of rotation close to 100 \kms]{corbelli2003, Saburova2012}.

The lower gas rotation speed found for the eastern side of the spiral galaxy Arp 68,  more distant from the satellite, may indicate either a large-scale non-circular motion of gas covering a significant part of the disc, or (more likely) that the stellar and gaseous discs, beyond 3--5~kpc from the center, do not lie in the same plane.

\section{Arp~58}\label{res2}
 
 \subsection{Kinematics of the emission gas}

\begin{figure}
%\hspace{-4.5cm}
\includegraphics[width=\linewidth]{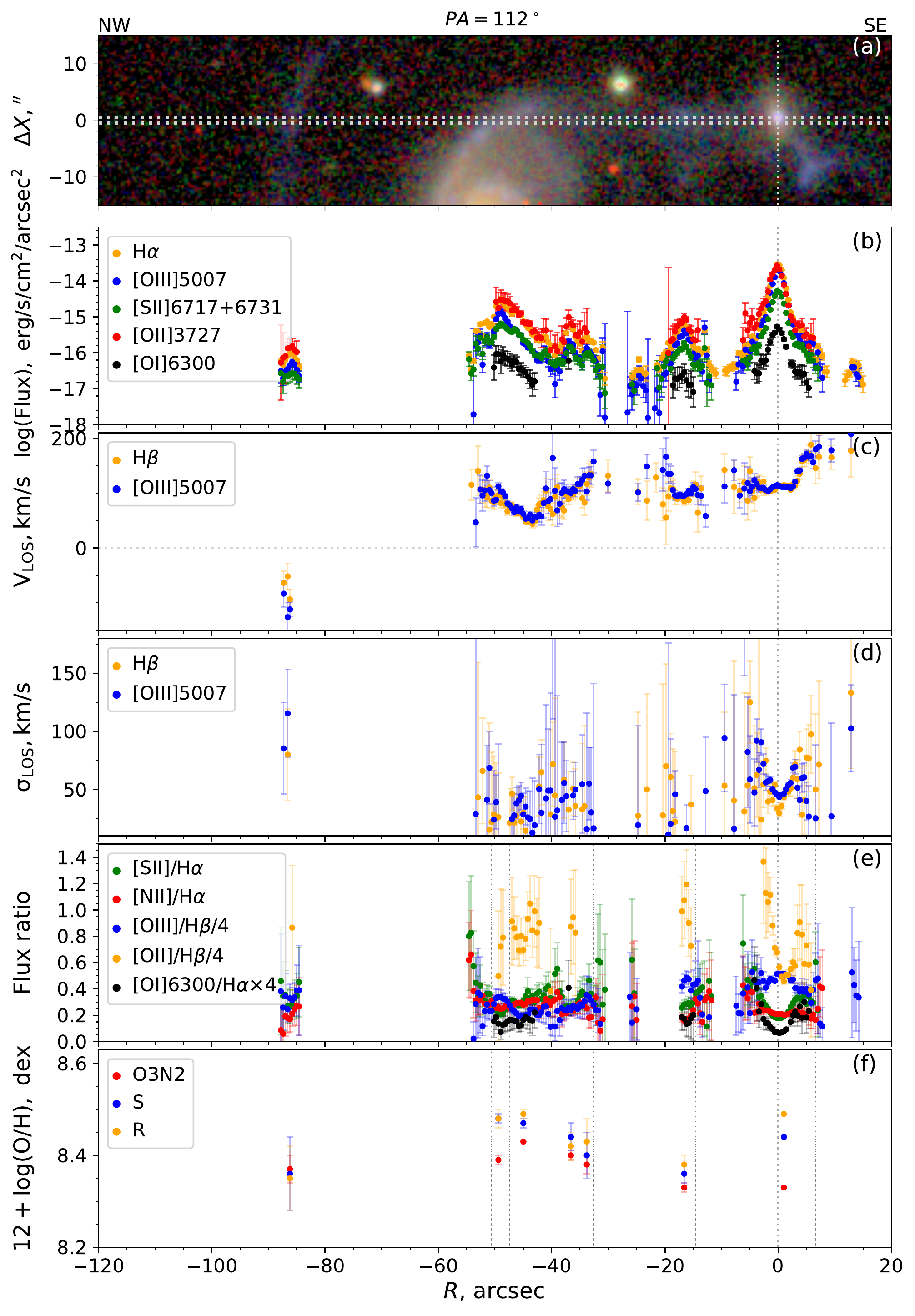}%11.5cm
\caption{The same as in Fig. \ref{arp68_pa95} but for Arp~58, $PA=112^\circ$. The value V=0 km s$^{-1}$ corresponds  to 11137 km s$^{-1}$.}
\label{arp58_pa112}
\end{figure}

\begin{figure}
%\hspace{-4.5cm}
\includegraphics[width=\linewidth]{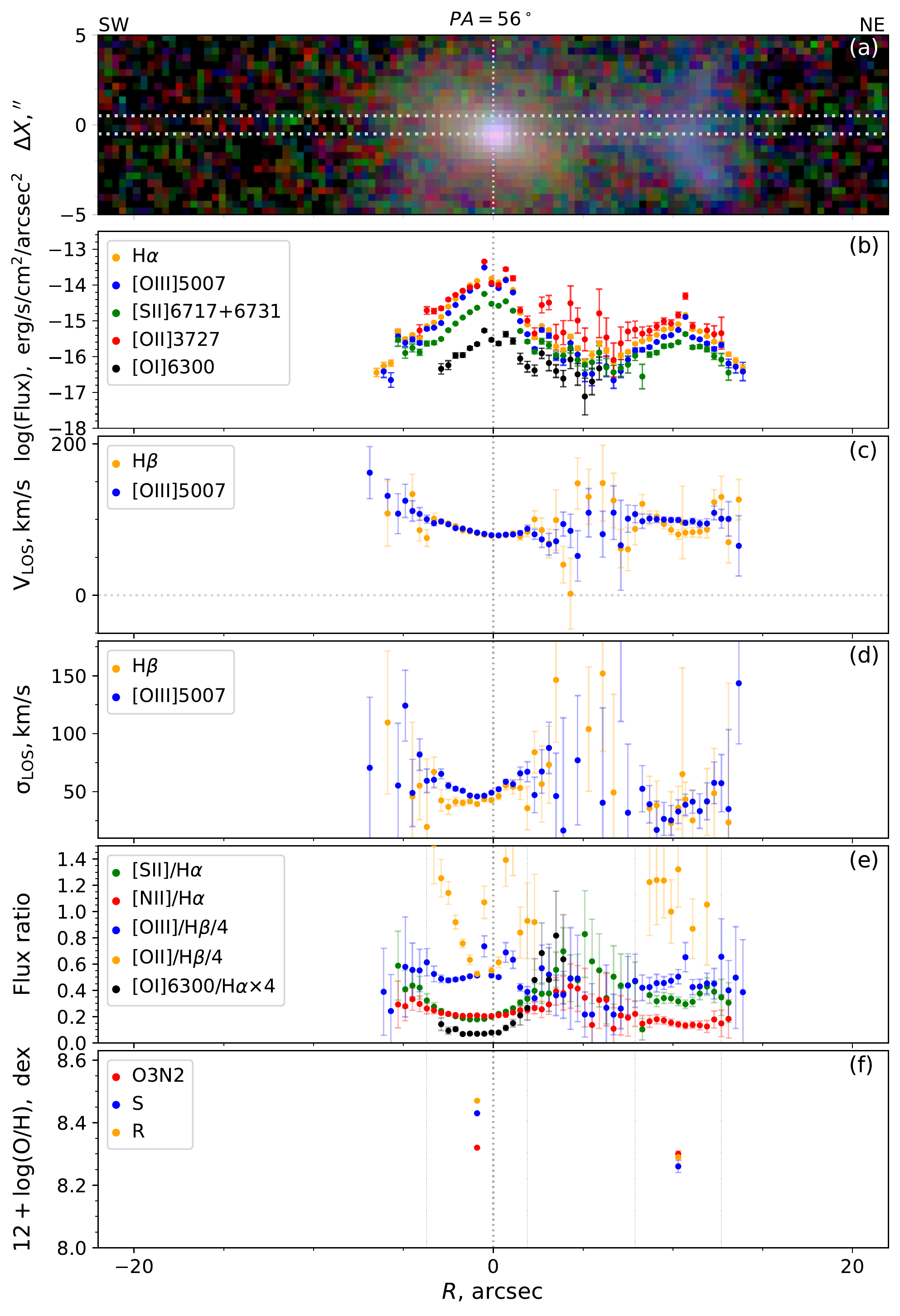}%11.5cm
\caption{The same as in Fig.\ref{arp68_pa95} but for Arp~58, $PA=56^\circ$. Zero velocity corresponds to 11137~\kms.}
\label{arp58_pa56}
\end{figure}

The variation of the measured values  along the slits for Arp~58 are shown in Figs. \ref{arp58_pa112}, \ref{arp58_pa56}. Following Paper~I, in this figure we adopted  the velocity of galaxy centre $V_0 = 11137$~\kms{} as the zero-point of LOS velocity, and the distance to the galaxy 160 Mpc, which  corresponds to 0.8~kpc per 1~arcsec. 

\begin{figure}
%\hspace{4.5cm}
\includegraphics[width=\linewidth]{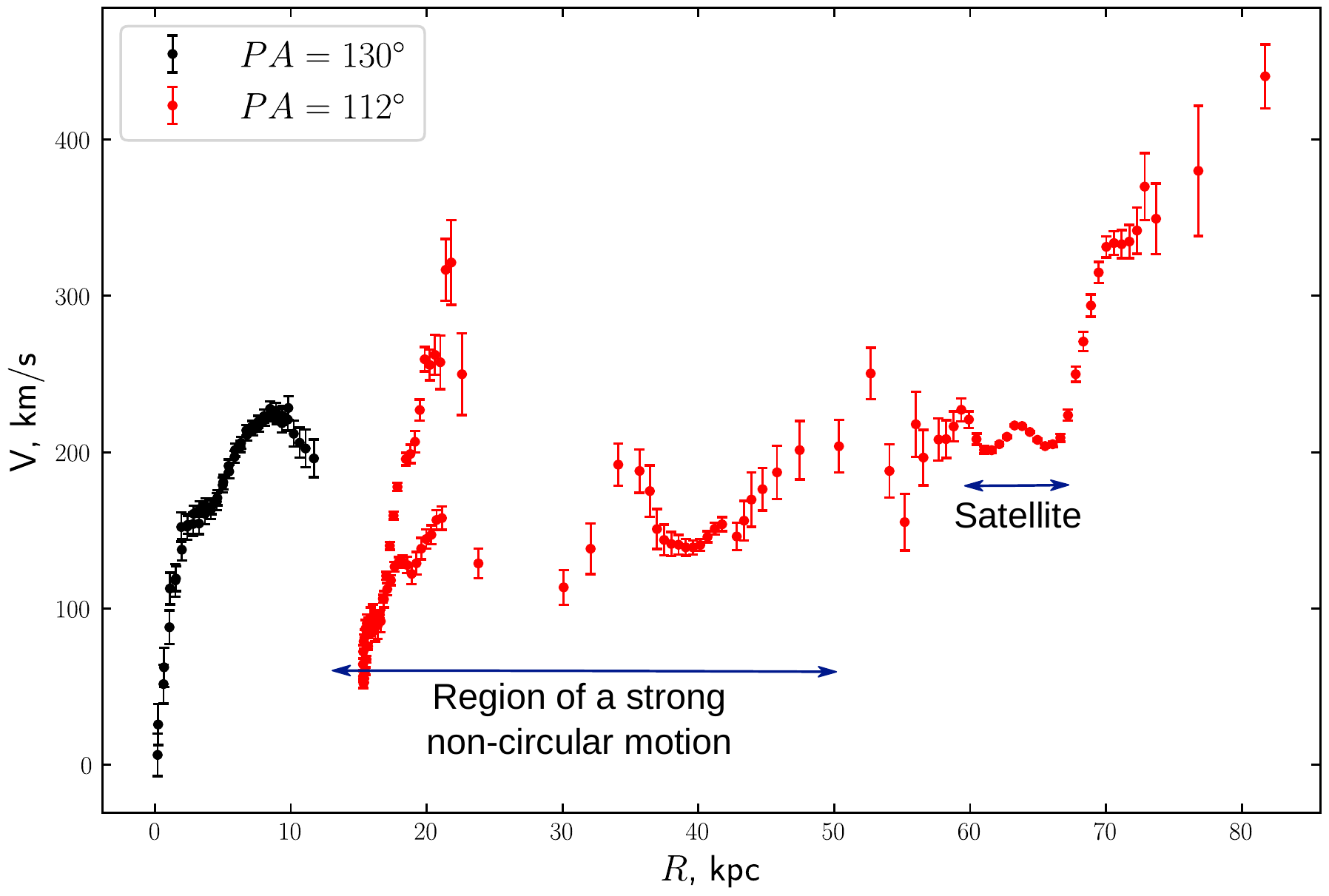}%11.5cm
\caption{A conditional rotation curve of Arp~58 corrected for projection effects assuming circular rotation of gas for two slit orientations. The data for PA=130$^\circ$ were taken from Paper I. For the slit at  PA=112$^\circ$, running along the spiral arm,  a strong non-circular motion of gas dominates in the extended region R = 13--50 kpc and  also beyond the satellite.}
\label{arp58_rc}
\end{figure}

As Fig. \ref{arp58_pa112} shows, the off-centre slit PA = 112$^\circ$ crosses both spiral arms, which have the opposite signs of LOS with respect to the centre. If the position angle PA$_0$ of kinematic major axis of this galaxy is close to  photometric value $PA_0\approx 164^\circ $ (HYPERLEDA), then the slit should intersect the minor axis  at the coordinate R $\approx -75$ arcsec in Fig. \ref{arp58_pa112}, in agreement with the expected zero velocity  in this region. Nevertheless the sparseness of points on the diagram in this region makes it difficult to estimate the kinematic minor axis precisely. Note that the velocities measured by [O~\textsc{iii}] line favour a slightly lower value of $PA_0$. We demonstrate the  conditional rotation curve of Arp~58 corrected for projection effects assuming circular rotation of gas for two slit orientations in Fig. \ref{arp58_rc}.

The most intriguing feature of the LOS profile at  PA = 112$^\circ$ is the deep minimum covering a wide range of distances  between  $R = -35$ and $-50$~arcsec.  It reveals a strong non-circular gas motion in this region, where the rotation of gas retards and its velocity component along the line of sight drops down at more than 50~\kms. This anomalous region shown in Fig. \ref{map2} by circle falls at the bright end of the spiral arm where it turns into a tidal arm, stretching towards the satellite. 

Note that such regions were predicted earlier for gas streams in the models of spiral interacting galaxies, appearing  at some stage of their dynamic evolution.   The perturbations of gas motion in the plane of a disc  caused by tidal forces lead to formation of regions there  where the gas orbits are intersecting  and the  dissipative gas streams are converging along narrow caustics due to orbit crowding. It results in the triggering the enhanced star formation in the local region of a spiral arm at the base of tidal tails called hinge clumps \citep{Hancock2009, StruckSmith2012,Smithetal2016}. Indeed, at least some galaxies with long curved tails clearly demonstrate the enhanced star formation expected for hinge clumps \citep{StruckSmith2012}. In our previous paper   a similar reduction of LOS velocity at about 30~\kms{} in the region of the hinge clump was found in the interacting galaxy NGC~4017 = Arp~305 \citep{Zasovetal2018}.

  The other noteworthy region crossed by the slit is the satellite and its stellar/gaseous extension. It is the actively star-forming dwarf galaxy, looking bright in the GALEX UV-image. It has  symmetric elliptical inner isophotes, while the  outer isophotes are distorted due to the interaction. 

Velocity  along the slit running at the angle of about 30$^\circ$ to the major axis of the satellite (PA = 56$^\circ$) shows a clear velocity gradient, which one can expect for rotating gaseous disc. To the SW from the center LOS velocity changes at about 30 km s$^{-1}$ at the distance r = 5 arcsec. Because the tilt angle of such a disc is unknown, one may provide only a rough lower estimate of the satellite's mass, assuming that the velocity gradient is due to rotation: $M_s > V_{LOS}^2r/G \approx 8\cdot10^8M_\odot$. Despite the very approximate  estimate, it fits nicely to the stellar mass of satellite $8.5\cdot10^8M_\odot$ found earlier  in Paper I.  

To the NE of the satellite, the slit PA = 56$^\circ$ passes through the bright UV extension (see the GALEX UV image in Paper I) which may be either a long (about 10 kpc) jet-like  tail of gas thrown out from the satellite, or the extended emission region at the end of the tidal spiral arm of the main galaxy projected onto the satellite and located in its vicinity. A  constant LOS velocity of gas and its low velocity dispersion along this  tail makes the latter version preferable. Its LOS velocity also agrees with the velocity of spiral arm  on the other side of the satellite.

\subsection{Gas-phase oxygen abundance in Arp~58}

\begin{figure*}
%\hspace{-4.5cm}
\includegraphics[width=\linewidth]{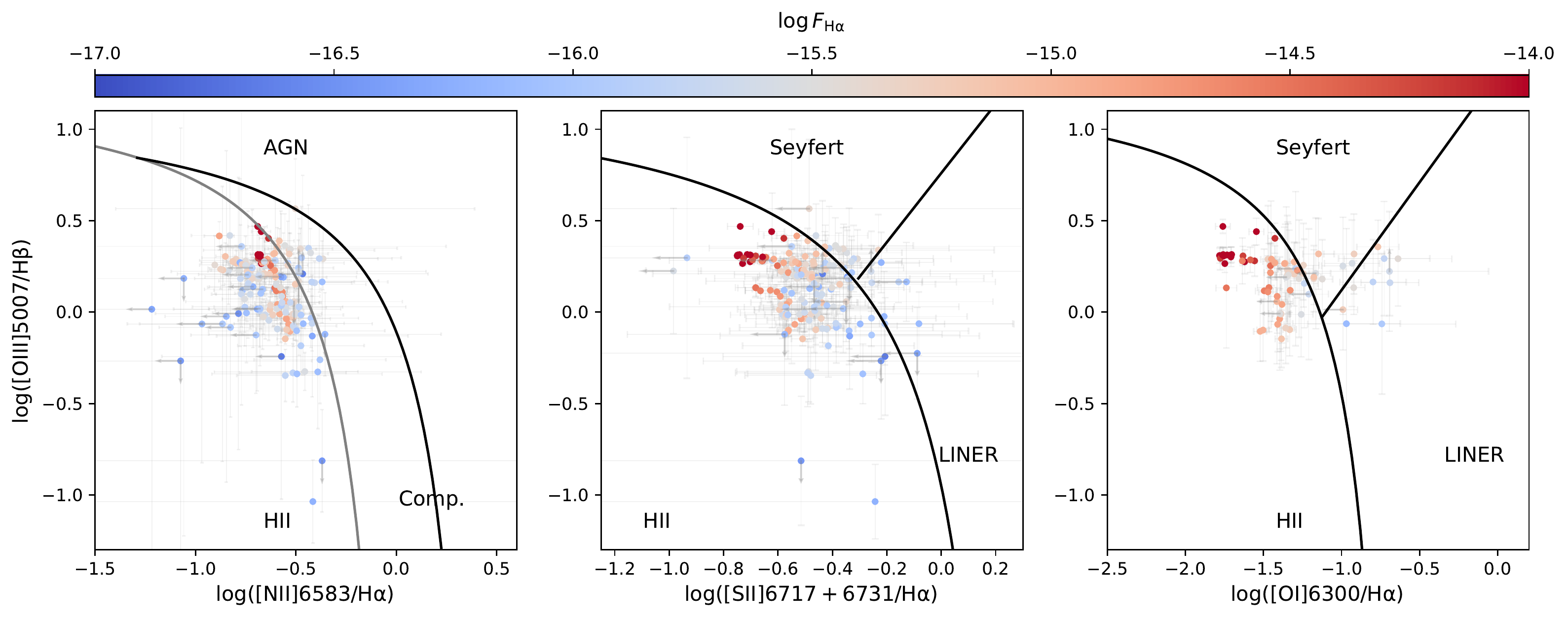}%11.5cm
\caption{Same as Fig.~\ref{arp68_bpt}, but for Arp~58.}
\label{arp58_bpt}
\end{figure*}

 The position of the emission regions of Arp~58, crossed by the slits, on the diagnostic diagrams, evidences (in most cases) the line ratios  expected for \HII{} regions (Fig.~\ref{arp58_bpt}). Ignoring a small number of points, which demonstrate  a significant role of non-photoionization mechanisms of excitation, we estimated the oxygen abundance in a same way as for Arp~68 (see Sec.~\ref{sec:Arp68_abund}, Figs \ref{arp58_pa112} and \ref{arp58_pa56}). In addition to O3N2 and S methods, we also used R-method \cite{Pilyugin16} based on the [O~\textsc{ii}] 3727,3729~\AA, H$\beta$, [O~\textsc{iii}] 4959, 5007~\AA{} and [N~\textsc{ii}] 6548, 6584~\AA{} line ratios as in case of Arp~58 we were able to measure all these lines simultaneously. 

 Three different methods we used are consistent with each other -- with the exception of the satellite, where the O3N2 method in both spectral cuts gives the value of (O/H) about 0.1~dex lower than the S and R methods. The reason of it is not obvious.  The region we considered is relatively bright, so it is unlikely due to insufficient S/N in some of the emission lines. In principle, the discrepancy between different oxygen abundance calibrators is the well known unsolved problem \citep{Kewley2008}, but this is the only bright \HII region in Arp~58 demonstrating such a disagreement. This region corresponds to the central nuclear part of the satellite where a strong radial gradient of the ionisation parameter is observed  (as follows from the relative distribution of [O~\textsc{ii}]/H$\beta$ and [O~\textsc{iii}]/H$\beta$ lines fluxes ratio, see panel (e) in Figs.~\ref{arp58_pa112},\ref{arp58_pa56}). This might be either related to the high contribution of the DIG in the disc of the satellite outside its central star cluster, or it may point out to the density bounded \HII region in the galaxy centre (which exhibits the excess of high-excitation lines like [O~\textsc{iii}] and might be associated with outflows), or rather both factors can contribute to the observed distribution of the emission lines fluxes. 
 
 Both the  presence of a central outflow or density bounded \HII regions may  lead to higher ratio of [O~\textsc{iii}] flux to that of lower excitation ions ([O~\textsc{ii}], [N~\textsc{ii}], [S~\textsc{ii}]) in this part of the galaxy in comparison with the normal \HII regions. In turn, it will lead to the underestimation of the metallicity by O3N2 method, which is proportional to $\log$([N\ \textsc{ii}]/[O\ \textsc{iii}]). In any case, the ionization parameter is significantly changing in this region, and thus IFU observations are necessary to clarify the ionization state and the metallicity of the region. Note, however, that $R$ and $S$ methods give $12+\log(\mathrm{O}/\mathrm{H})\approx8.4$--8.5, which is quite expected for galaxies with a stellar mass of about $10^9M_\odot$.

\begin{figure}
%\hspace{-4.5cm}
\includegraphics[width=\linewidth]{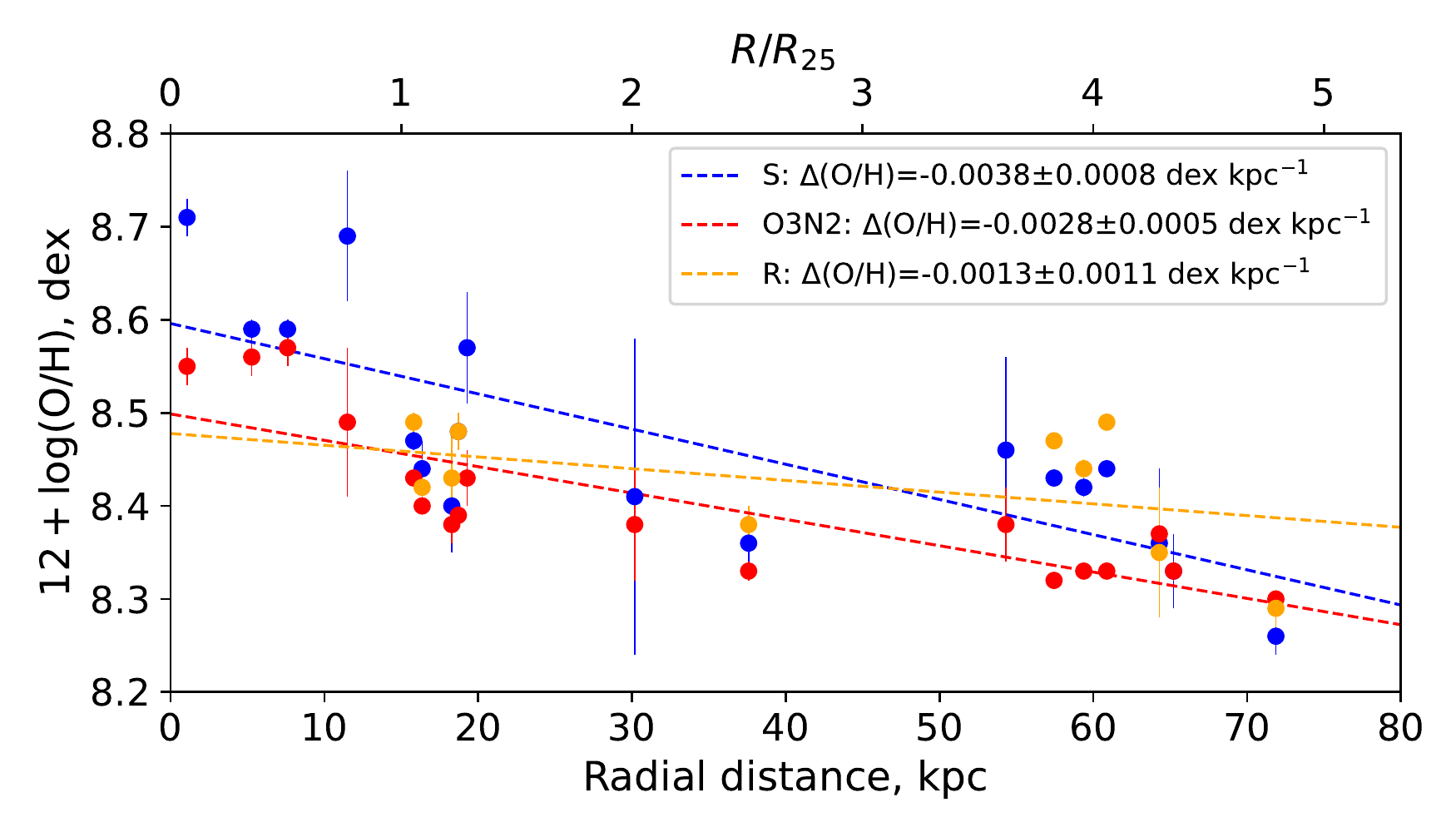}%11.5cm
\caption{The radial (deprojected) distribution of the oxygen abundance for  Arp~58.}
\label{oh}
\end{figure}

A general trend of the abundance variation with the radial distance in the main galaxy is evident from $PA=112^\circ$ data: the maximum of ($\mathrm{O}/\mathrm{H}$) corresponds to the region closest to the center. The radial gradient  of the oxygen abundance (Fig.~\ref{oh}) agrees with the data found for this galaxy in Paper~I.

\begin{figure}
%\hspace{-4.5cm}
\includegraphics[width=1.15\linewidth]{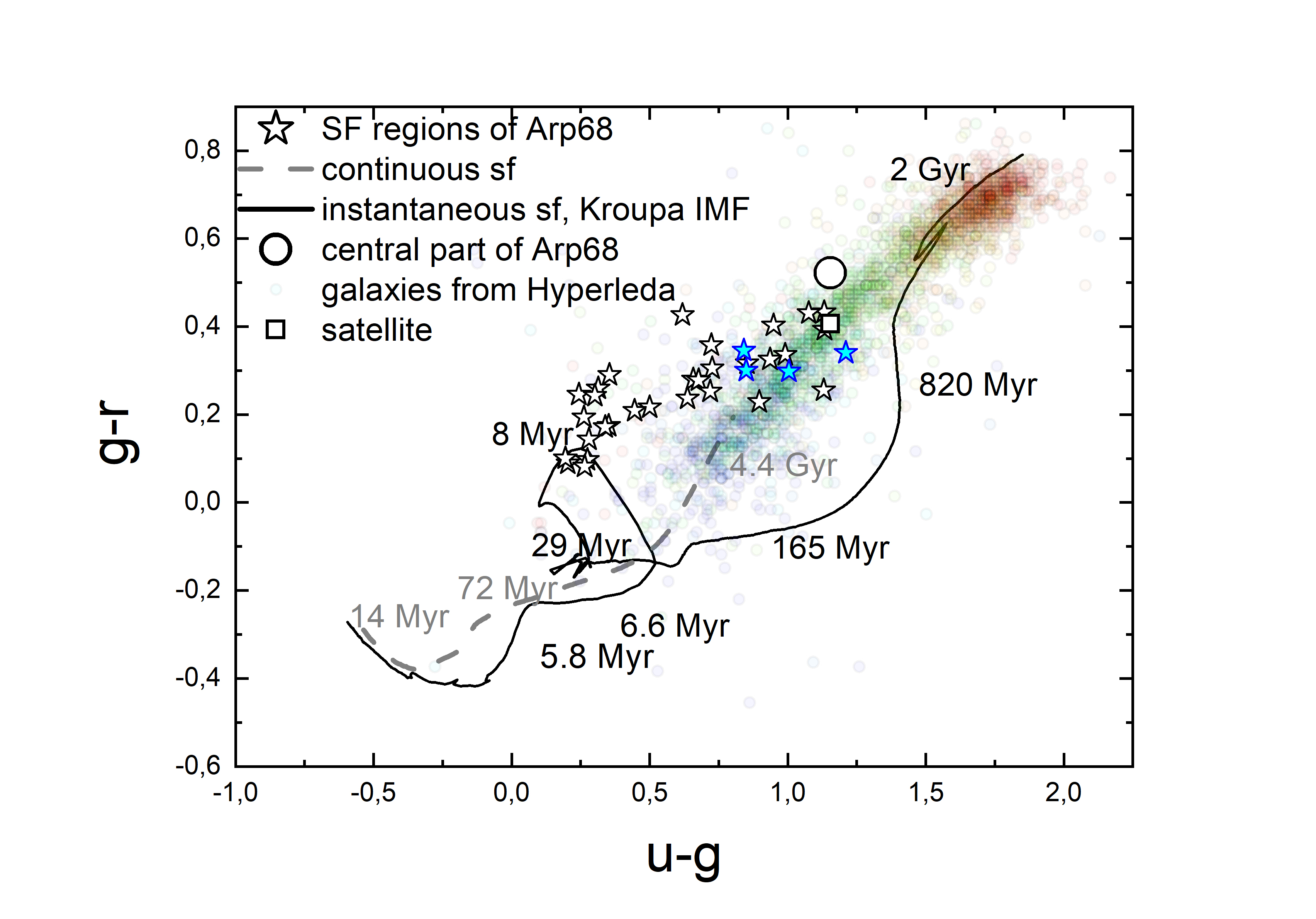}%11.5cm
\caption{The (g-r) vs (u-g) diagram plotted for the  star-forming regions in spiral arms of Arp~68 (open stars) in comparison with the central regions of the spiral galaxy (large open circle) and the satellite (open square). Blue asterisks mark the regions located  on the periphery of the disc.  For comparison,  the colour sequence for bright galaxies of different morphological types is shown (see the text).  
Black line and gray dashed line with the signed ages are the  Starburst99 model tracks for instantaneous and continuous star formation respectively.}
\label{colors_arp68}
\end{figure}

\begin{figure}
%\hspace{-4.5cm}
\includegraphics[width=1.15\linewidth]{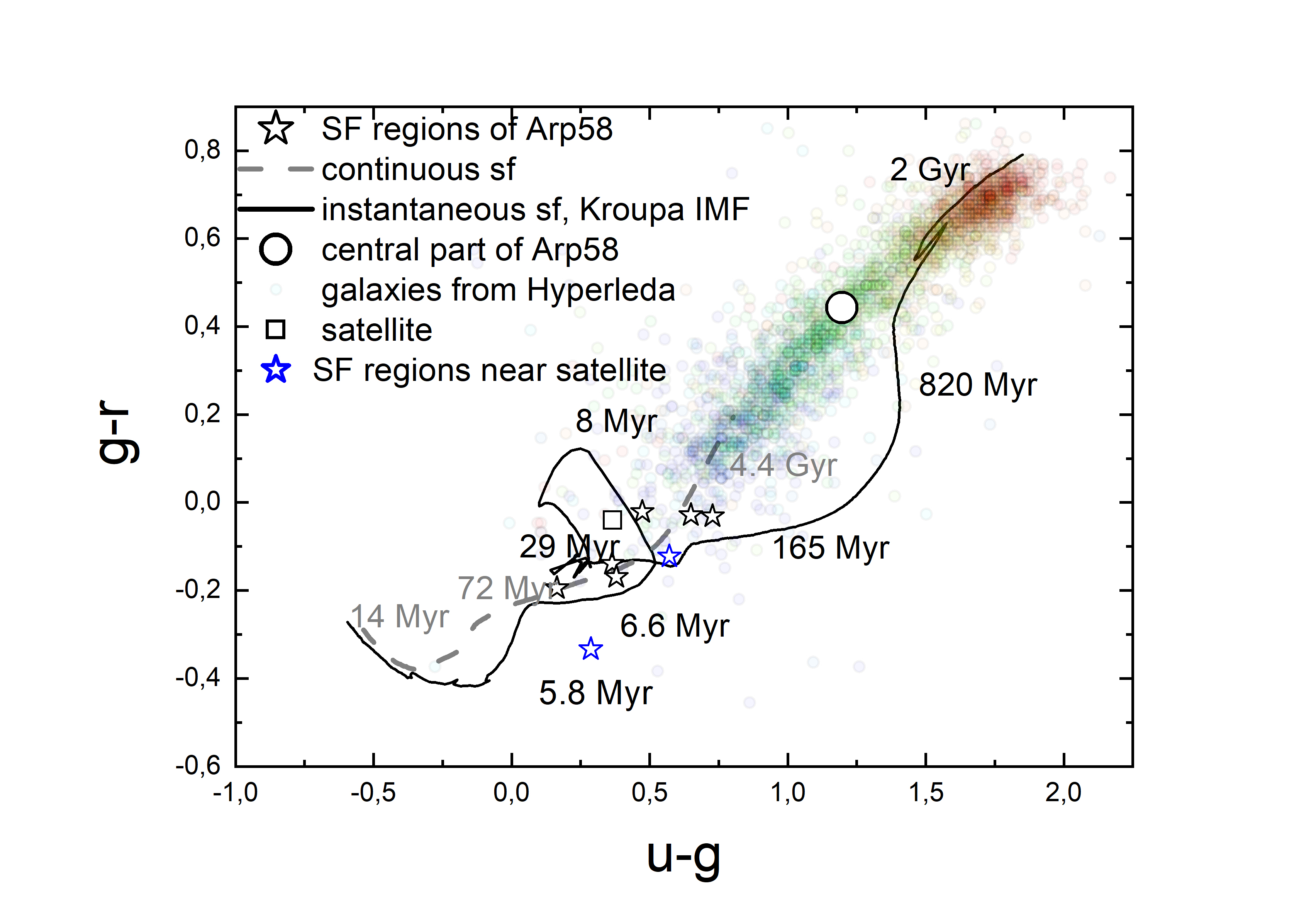}%11.5cm
\caption{The same two-colour diagram as in Fig. \ref{colors_arp68} but for Arp~58. Blue stars show the positions of star-forming regions near the satellite.}
\label{colors_arp58}
\end{figure}

\section{Discussion}\label{sec:disc}
Evidently, we observe  both spiral galaxies described in this work at a similar dynamical stage of close interaction with the nearby satellites. At this stage, after the  last close encounter,  a tidal arm has grown and stretched towards the intruder. This process, leading to the formation of a system of type M~51, has developed  before the satellite  run too far away from the spiral galaxy or  sank into its  stellar disc as the result of merging. 

It is worth comparing the systems  we study.  Table \ref{comparison} gives the properties of the  systems.  Masses of \HI were derived from the Hyperleda database parameter $m_{21,c}$ for Arp~58 and from \citet{Haynes2018} for Arp~68,  FIR data were taken from \cite{Casasola2004}. To get the rate of star formation $SFR_{FIR}$ we used the linear relationship between SFR and FIR, following \cite{Iglesias2006}. The stellar mass $M_*$  of the satellite for Arp~58 is taken from Paper I. For the satellite of Arp~68 we calculated $M_*$  from the {\it g}-band luminosity obtained from the SDSS image processing and the stellar mass-to-light ratio based on the $(g-r)$ index using  the model relation `M/L -- colour' from \cite{Roediger2015}. We also calculated star formation rate from the  $FUV$-band GALEX  data, following   \citep{Wyder2009}, which used the relation between SFR and UV flux from \citep{Kennicutt1998}. 
\begin{table}
\caption{Comparison of the properties of Arp~58 and Arp~68.} \label{comparison}
\begin{center}
\begin{tabular}{ccc}
\hline\hline
	&Arp~68&	Arp~58\\
	\hline
$M_*/M_\odot$&	$6.1\cdot 10^9$&	$9 \cdot 10^{10}$\\
$(M_{*}/M_\odot)_{sat}$&	$2.4\cdot 10^8$&	$8.5\cdot10^8$\\
$\Delta V_r$&	$\sim 40$~\kms & $\sim 60$~\kms\\
Vmax	&100--130~\kms& 	230--250~\kms\\
$M_{\rm HI}/M_\odot$&	$6.5 \cdot10^9$&	$3.6\cdot10^{10}$\\
$M_{\rm gas}/M_*$&	1.44&	0.54\\
grad (O/H)&	<0.03&	0.003\\
$L_{FIR}/M_{\rm HI}$&	0.84&	1.34\\
$SFR_{FIR}$&	0.7 $M_\odot/yr$&	4.8 $M\odot/yr$\\
$SFR_{FUV}$&	1.5$\pm0.08$ $M_\odot/yr$&	3.5$\pm0.15$ $M\odot/yr$\\
\hline
\end{tabular}
\end{center}
\end{table}

The most significant difference between the two systems lies in their masses:  Arp~68 is more than an order of magnitude lighter by stellar mass than the more distant   galaxy Arp~58. The difference in the total  masses of galaxies is also evidenced by the twice higher rotation velocity of the latter. Stellar masses of the companions in both systems are tens times lower than that of the main galaxies. 

However, in spite of the big difference of their masses, many characteristics of the systems are very similar. The LOS  velocities of the satellites differ little from the central velocities of the main galaxies  (see the velocity difference $\Delta V_r$ in Table \ref{comparison}), and are close to the velocities of the regions of spiral arms adjacent to satellites, which perfectly agrees with the expectation that the orbital rotation of satellites is in the same direction as the rotation of galaxies. Indeed, the latter  is common for M~51-type systems \citep{Reshetnikov2003}. A low radial oxygen abundance gradient grad (O/H) we found for Arp~58 is also typical for interacting galaxies. 

In the case of Arp~68  our measurements cover too narrow range of radial distances to estimate the gradient reliably. However it demonstrates the reduced oxygen abundance, which may be caused by the position of this system inside of the Eridanus void (see Section \ref{sec:Arp68_abund}). 

Note that the ratio of mass of gas $M_{\rm gas}\approx 1.35M_{\rm HI}$ to the stellar mass $M_*$ ($M_{\rm gas}/M_*$ in Table \ref{comparison}) shows that in both galaxies a gas layer makes a high contribution to the  mass of the disc component. A total mass of gas in Arp~68 is of the same order as the stellar mass of this  spiral galaxy.\footnote{ Here we compare the stellar mass of spiral galaxies with the total  mass of HI of the systems,  ignoring that the single dish estimate of HI also includes a gas containing  in small satellites. However even if we add the expected stellar mass of the satellites to $M_*$ to compare it with the total mass $M_{\rm HI}$, the result will remain the same.} It may explain the presence of a well ordered  spiral pattern in Arp~68, which is rarely observed in galaxies for which the velocity of disc rotation does not exceed 100~\kms. As numerical simulations show, a formation of spiral wave pattern in slowly rotating galaxies requires special conditions although it is facilitated in the case of a large relative gas mass \citep{Zasov2021}. A tidal force is the additional factor which causes a contrast two-armed structure with the intense star formation in the local regions along the tidal arm.

It is worth comparing the total star formation rate in  the spiral galaxies we consider. Both  are the actively star-forming systems.  It follows from the strong emission lines in their spectra and the observed blue clumps of young stellar population,   most prominent in  the spiral arms of Arp~68.  Although star formation rates (SFR) of spiral galaxies Arp~68 and Arp~58 obtained from FIR luminosities are different (see Table \ref{comparison}),  in both cases they  agree with the main sequence for star-forming galaxies at the diagram `SFR vs $M_*$'  \citep[see e.g.  ][]{Pearson2018,Chang2015}.   
The efficiencies of the star formation (${\rm SFE} = {\rm SFR}/M_{\rm HI}$) of these galaxies are also similar  and quite normal for spiral galaxies $(\log {\rm SFE}  =  1.5\cdot10^{-10}yr^{-1} $ for Arp~68 and $1.3\cdot10^{-10}yr^{-1} $ for Arp~58).  It  shows that both galaxies have evolved most of their lives as the usual  late-type  galaxies.  However, in Arp~68 situation is more complicated:  the approximately twice higher integral star formation rate found from UV radiation in comparison with FIR data in this spiral galaxy indicates that in the modern era its star formation rate  is somewhat higher than the average. It agrees with the intensity of its emission spectrum: the current time  SFR, found from $H_\alpha$ narrow band photometry  \citep{James2004}, re-scaled to the accepted distance, also gives a higher value: SFR =  3.8 $\pm$0.7 $M_\odot/yr$.   A low consumption time of \HI,  defined as $\tau_{\rm HI} = M_{\rm HI}/{\rm SFR}$, is close to one billion yr, which is more typical for star-forming  molecular clouds.  It gives evidence that this galaxy experiences a mild burst of star formation in the current time, evidently caused by the interaction with its very close satellite crossing the peripheral regions of the galaxy.
 
 In Figs. \ref{colors_arp68}, \ref{colors_arp58} we show the positions of the star-forming regions of Arp~68 and Arp~58 on the $(g-r)$ vs $(u-g)$ colour diagram (open asterisks) obtained by processing of SDSS images within the aperture which diameter depends on the visible size of a knot (in most cases between 1 and 3 kpc).  Most of these regions are located in the spiral arms of galaxies. Small transparent circles trace  the normal colour sequence for galaxies of different morphological types constructed 	on the base of Hyperleda data for about 2.5 thousands  of bright galaxies. Elliptical  galaxies occupy the upper part of the sequence, while irregular galaxies with active star formation dominate at the bottom. 
 The large open circle and the open square in both diagrams  illustrate the colour  of the central parts of  spiral galaxies and  their satellite respectively. 
 Continuous and  dashed lines  follow  Starburst99 model, which tracks the evolution of colour for instantaneous and continuous star formation respectively \citep{Leitherer1999}. We mark the model ages of stellar population along the tracks.
 
 As it follows from Fig. \ref{colors_arp68}, the  colours of the central part of the galaxy and its  satellite well agree with the normal colour sequence. They occupy the position of normal galaxies with the moderate SFR.  The other regions of the disc we consider  form the chain of asterisks, corresponding to the mixture of young stars with the usual stellar population of late-type galaxies taken in different proportions.  They could be conditionally divided into two groups on the diagram: one group is formed by the relatively ''red'' faint regions, some of them are located at the periphery with low surface density (blue asterisks). They contain a mixture of stellar population of different ages with the stellar composition close to that of Sc-Sb type galaxies. It may be considered as the evidence of a long-lasting or relatively low intense star formation at the background of the older stellar population in these local regions. The second group presents the bright regions of recently formed stars. They  lay close to the evolutionary track for instantaneous star formation with the ages of about 8~Myr.  Most of these recently formed  regions belong to the spiral arm of Arp~68,  which is extended toward the satellite.
 
 For Arp~58 the picture is different. Both spiral arms demonstrate the presence of  a young stellar population with the ages of several dozens of Myr  without significant contribution of older stars to the luminosity. The position of the clumps near the satellite is marked in the diagram (Fig. \ref{colors_arp58}) by blue colour, the other star-forming sites are related to spiral arms.  In all these regions  a recently formed stellar population prevails. There we observe the regions  of star formation against a background of a much older stellar population.  However the total star formation rate of this galaxy is normal for galaxies with a given content of gas in a disc, so  the influence of the satellite on star formation  is not evident.

 The satellites of spiral  galaxies we consider are also different: in Arp~58 it is bright in the UV GALEX images (see Paper I),  while in Arp~68 the satellite is very dim in UV, and its optical colour evidences the  moderate star formation  with a large contribution of the old stellar population to its luminosity.

 We also tried to find the properties of stellar population in  the regions chosen for color-color diagrams by applying the more sofisticated method of wide-range SED plus spectral fitting.\footnote{ We used the {\sc nbursts+phot} technique, described in detail in \citet{ChilingarianKatkov2012}} To construct the SED for a wide range of wavelengths  we used the images taken in the databases of  GALEX (FUV, NUV), SDSS  (u, g, r, i) and DECaLS z-bands, parallel with the  data of the Spitzer Space Telescope for Arp 68 and the WISE data for Arp 58.   The results of SED+ optical spectra fitting led qualitatively to similar conclusions concerning  stellar ages of chosen regions to those obtained by color-color analysis. 
 In paticular, we obtained the young ages of stellar population for the spiral arms in both spiral galaxies lying  between several dozens and 100 Myr; the central part of Arp~58 is well described by the mixture of old ($T_1\sim 5$ Gyr) and young  ($T_2 \sim 30$  Myr) stellar components in agreement with our previous findings. In Arp 68 both the central region of spiral galaxy and its satellite contains a comparable input of young  and medium age stellar population with $T < 10^9$ yr.  More details on the SED+spectra fitting for interacting systems, including those considered above, and the discussion of results, will be the item of forthcoming papers.

\section{Conclusions}\label{conclusions}

Summarizing, we conclude that despite the large difference in  stellar masses of the  considered M~51-type galaxies, their interaction with a close companion has common features. In both cases, the observed velocities of the satellites are close to the velocities of the disc regions adjacent to them, which promotes to strong tidal disturbances. In addition to the asymmetry of radial LOS  distribution (at least in Arp~68), both spiral galaxies reveal a strong  non-circular motion of gas both in the local kpc-scale and in the more extended regions of their  discs which is not often seen  in other galaxies of M~51- type (see Sect. \ref{intro}).  

The question remains open whether these disturbed regions  may be accounted for a local bursts of star formation induced by the  tidal interaction with the satellite or there are some other reasons for gas perturbation such as the accretion of halo gas onto a disc induced by the interaction. 
 The total star formation rate in the spiral galaxies under consideration does not indicate the  powerful large-scale induced outbursts, although its current value in the Arp 68 spiral galaxy is apparently somewhat increased.

It seems that both galaxies we consider  are observed in a similar phase of  tidal interaction with nearby satellites, accompanied by the development of regular two-armed structure and by strong local perturbations in gas velocities. Nevertheless,  galaxies differ by star formation rate  and by chemical evolution of gas, which is evidently associated with the big difference in their masses and the relative masses of gas in their discs. 

In the case of Arp~68, the gas-phase oxygen abundance is lower than expected for its luminosity  probably being affected by the low density environment by analogy to the other low-luminous galaxies located in the voids (see Section \ref{sec:Arp68_abund}).

\section*{Acknowledgements} 
The authors deeply thank the referee for the comments and discussion.
AS research was supported by The Russian Science Foundation (RSCF) grants No. 19--12--00281 (spectral analysis) and 19--72--20089 (photometric analysis). 
Part of the observational data was obtained on the unique scientific facility the Big Telescope Alt-azimuthal SAO RAS and the data processing 
was supported  under  the   Ministry of Science and Higher Education of the Russian Federation grant  075--15--2022--262 (13.MNPMU.21.0003). 
OE acknowledge funding from the German Research Foundation (DFG) in the form of an Emmy Noether Research Group (grant number KR4598/2--1, PI Kreckel).

The authors acknowledge the usage of the
HyperLeda database (http://leda.univ-lyon1.fr). 
The work is partly based on the images from the DECaLS and SDSS surveys. 
The Legacy Surveys consist of three individual and complementary projects: the Dark Energy Camera Legacy Survey (DECaLS; NOAO Proposal ID  2014B-0404; PIs: David Schlegel and Arjun Dey), the Beijing-Arizona Sky Survey (BASS; NOAO Proposal ID 2015A-0801; PIs: Zhou Xu and Xiaohui Fan), and the Mayall z-band Legacy Survey (MzLS; NOAO Proposal ID  2016A-0453; PI: Arjun Dey). DECaLS, BASS and MzLS together include data obtained, respectively, at the Blanco telescope, Cerro Tololo Inter-American Observatory, National Optical Astronomy Observatory (NOAO); the Bok telescope, Steward Observatory, University of Arizona; and the Mayall telescope, Kitt Peak National Observatory, NOAO. The Legacy Surveys project is honored to be permitted to conduct astronomical research on Iolkam Du'ag (Kitt Peak), a mountain with particular significance to the Tohono O'odham Nation.

Funding for the Sloan Digital Sky Survey IV has been provided by the Alfred P. Sloan Foundation, the U.S. Department of Energy Office of Science, and the Participating Institutions. SDSS-IV acknowledges
support and resources from the Center for High-Performance Computing at
the University of Utah. The SDSS web site is www.sdss.org.

SDSS-IV is managed by the Astrophysical Research Consortium for the 
Participating Institutions of the SDSS Collaboration including the 
Brazilian Participation Group, the Carnegie Institution for Science, 
Carnegie Mellon University, the Chilean Participation Group, the French Participation Group, Harvard-Smithsonian Center for Astrophysics, 
Instituto de Astrof\'isica de Canarias, The Johns Hopkins University, 
Kavli Institute for the Physics and Mathematics of the Universe (IPMU) / 
University of Tokyo, Lawrence Berkeley National Laboratory, 
Leibniz Institut f\"ur Astrophysik Potsdam (AIP),  
Max-Planck-Institut f\"ur Astronomie (MPIA Heidelberg), 
Max-Planck-Institut f\"ur Astrophysik (MPA Garching), 
Max-Planck-Institut f\"ur Extraterrestrische Physik (MPE), 
National Astronomical Observatories of China, New Mexico State University, 
New York University, University of Notre Dame, 
Observat\'ario Nacional / MCTI, The Ohio State University, 
Pennsylvania State University, Shanghai Astronomical Observatory, 
United Kingdom Participation Group,
Universidad Nacional Aut\'onoma de M\'exico, University of Arizona, 
University of Colorado Boulder, University of Oxford, University of Portsmouth, 
University of Utah, University of Virginia, University of Washington, University of Wisconsin, 
Vanderbilt University, and Yale University.

 \section*{DATA AVAILABILITY}
The data underlying this article will be shared on reasonable request to the corresponding author.

\bibliographystyle{mnras}
\bibliography{arp68}
\label{lastpage}
\end{document}